\documentclass[epsf]{aastex}
\usepackage{apjfonts}
\usepackage[onecolumn]{emulateapj5}
\shorttitle{Neutrino cooled accretion disks}

\shortauthors{Lee, Ramirez-Ruiz \& Page}

\begin{document}

\title{Dynamical evolution of neutrino--cooled accretion disks:
detailed microphysics, lepton-driven convection, and global
energetics}

\author{William H. Lee\altaffilmark{1}, Enrico
Ramirez--Ruiz\altaffilmark{2,3} and Dany Page\altaffilmark{1}}

\altaffiltext{1}{Instituto de Astronom\'{\i}a, Universidad Nacional
Aut\'{o}noma de M\'{e}xico, \\ Apdo. Postal 70-264, Cd. Universitaria,
M\'{e}xico D.F. 04510}

\altaffiltext{2}{School of Natural Sciences, Institute for Advanced
Study, Princeton, NJ, 08540}
\altaffiltext{3}{Chandra Fellow}

\begin{abstract}
We present a detailed, two dimensional numerical study of the
microphysical conditions and dynamical evolution of accretion disks
around black holes when neutrino emission is the main source of
cooling. Such structures are likely to form after the gravitational
collapse of massive rotating stellar cores, or the coalescence of two
compact objects in a binary (e.g., the Hulse--Taylor system). The
physical composition is determined self consistently by considering
two regimes: neutrino--opaque and neutrino--transparent, with a
detailed equation of state which takes into account neutronization,
nuclear statistical equilibrium of a gas of free nucleons and alpha
particles, blackbody radiation and a relativistic Fermi gas of
arbitrary degeneracy. Various neutrino emission processes are
considered, with $e^{\pm}$ capture onto free nucleons providing the
dominant contribution to the cooling rate. We find that important
temporal and spatial scales, related to the optically thin/optically
thick transition are present in the disk, and manifest themselves
clearly in the energy output in neutrinos. This transition produces an
inversion of the lepton gradient in the innermost regions of the flow
which drives convective motions, and affects the density and disk
scale height radial profiles. The electron fraction remains low in the
region close to the black hole, and if preserved in an outflow, could
give rise to heavy element nucleosynthesis. Our specific initial
conditions arise from the binary merger context, and so we explore the
implications of our results for the production of gamma ray bursts.
\end{abstract}


\section{Introduction}\label{intro}

Gas accretion by a concentration of mass is an efficient way to
transform gravitational binding energy into radiation \citep{s64,z64},
and is responsible for observable phenomena at all scales in
astrophysics. The more compact the object, the greater the
efficiency. In many cases, systems are observed in steady or
quasi--steady state during this process (e.g, CVs, LMXBs and
AGNs). Some external agent (e.g., the interstellar medium, a companion
star) provides a continuous supply of mass, energy and angular
momentum over a timescale that is much longer than the accretion
time. The energy dissipated by the flow before it is accreted, and
more importantly, what happens to this energy, determines many of the
properties of the flow itself. In most cases cooling is present
through some radiative electromagnetic process, which acts as a sink,
removing the dissipated energy. In the standard thin--disk theory
developed by \citet{ss73}, it is extremely efficient, maintaining a
``cool'' (in the sense that $kT \ll GMm_{p}/r$, where $M$ is the mass
of the central object) and thin (with a scale height $H \ll r$),
nearly Keplerian disk. The requirements of steady state, hydrostatic
equilibrium in the vertical direction, energy and angular momentum
balance, and an equation of state yield a solution for all relevant
variables as a function of the disk radius, $r$. The key question of
angular momentum transport was addressed by Shakura \& Sunyaev with
their famous $\alpha$ prescription, which allows a parametrization of
the viscous stresses and energy dissipation rates. Magnetic fields
could be at the physical origin this effect, through the
magneto--rotational instability \citep[MRI,][]{bh91,hb91}.  In the
case of inefficient cooling, a solution was found in the 1970s for
so--called slim disks \citep{sle76}, and later for a class that became
known as ADAFs \citep[Advection Dominated Accretion
Flows,][]{i77,ny94,ny95,acklr95}. In this case, the cooling is
negligible, either because the optical depth is large, or because the
density is so low that the radiative efficiency is extremely
small. The dissipated energy is advected with the flow and may be
absorbed by the central object. Such flows are geometrically thick,
with scale heights $H \simeq r$. What little radiation does emerge
from them arises, as in the thin disks, through various
electromagnetic processes.

In either of the above scenarios, accretion (and the accompanying
radiation) is usually thought to be limited by the Eddington rate, a
self--regulatory balance imposed by Newtonian gravity and radiation
pressure. The standard argument gives a maximum luminosity
$L_{\rm Edd}=1.3 \times 10^{38} (M/M_{\sun})$~erg~s$^{-1}$. Although 
this
may not be strictly the case in reality --- as in the current argument
concerning the nature of ULXs observationally \citep{rpp05,kd05}, and
also quite general theoretical considerations \citep{a05} --- it does
exhibit the qualitative nature of the effect of radiation pressure on
accreting plasma, in the limit of large optical depth.

The problem is circumvented (or at least deferred by nearly sixteen
orders of magnitude in luminosity) if the main cooling agent is
emission of neutrinos, instead of photons. This regime requires
correspondingly large accretion rates, of the order of one solar mass
per second, and is termed hypercritical accretion \citep{c89}. It is
important, for example, in the context of post--supernova fallback
accretion onto a proto--neutron star. In such a situation, the
densities and temperatures are so large ($\rho \simeq
10^{12}$~g~cm$^{-3}$, $T\simeq 10^{11}$~K) that photons are completely
trapped, and energetic neutrinos are emitted in large amounts, cooling
the gas and allowing accretion to proceed. The reader may wish to
consult the very clear introduction in this context given by
\citet{hc91}. When the stellar envelope experiences fallback onto the
central object, the system may be considered to be nearly in a steady
state, because of the timescales involved. There are other, more
violent situations, relevant to the study of gamma-ray bursts (GRBs)
in which the assumption of steady state is not justified because of a
time--varying mass and energy supply, and which therefore require a
dynamical analysis for their proper description.

In general, whether the system cools via electromagnetic radiation or
neutrinos, analytic steady state solutions are found by assuming that
the cooling is either efficient (thin disk) or inefficient (thick
disk). Over the last several years, many groups have considered these
cases in the neutrino cooled regime
\citep{pwf99,npk01,km02,dpn02,yua04} and established the general
features of the solutions. The real solution may be very different, in
particular (and in what concerns us for the following) because of the
previous history of the gas that constituted the accretion disk and
how the disk formed. It turns out this may be quite important in the
case of GRBs. Numerical analysis of the evolution of the disk with no
assumptions concerning the steady state has been carried out by
\citet{srj04}. They have reported a three dimensional calculation with
detailed microphysics. Unfortunately, computational limitations do not
allow one to continue such calculations for more than approximately
50~ms, whereas the typical short GRB lasts 0.2~s.

The sources of GRBs are now established (at least based on
observations of X--ray, optical and radio counterparts) to lie at
cosmological distances, with redshifts $z\simeq 1-4$ \citep[][and
references therein]{vkw00}. At such scales, the absolute energetics of
each event is approximately $10^{52}$~erg, assuming isotropic
emission. Apparent collimation inferred from achromatic breaks in the
afterglow light curve may reduce this to $10^{50}$~erg, depending on
the particular event \citep{fetal01,pk01,bkf03}. Although one
generally considers two classes of GRBs based on duration
\citep[shorter and longer than about 2~s,][]{ketal93}, all afterglows
to date (and the corresponding inferences) come from long bursts
\citep[although see][]{lrg01}. Strong evidence in favor of a SN/GRB
association \citep{ketal98,setal03,hjorth03} now comes from several
events (albeit all relatively nearby), hinting at an underlying
physical connection between the two phenomena. This is the central
basis of the collapsar model \citep{w93,mw99}, in which rotation of
the pre--supernova star allows for the creation of a massive accretion
disk, fed by the infalling stellar envelope. The GRB is then powered
for a fallback time, which can be long enough to account for the
observed durations. In the case of short bursts there is less evidence
to go on, but one possibility is that they arise from compact binary
mergers, with various combinations of black holes and neutron stars in
the binary \citep{ls76,bp86,elps89,bp91,nara92} such as in the first
binary pulsar to be discovered \citep{ht75}. The tidal disruption of
one star by the other gives rise to an accretion disk which powers the
GRB. A black hole is either present from the start (as a variation of
the Hulse--Taylor system) or is produced by the collapse of a
supramassive neutron star shortly after the merger itself. In this
case there is no external agent feeding the accretion disk, and thus
the event is over roughly on an accretion timescale (which would be on
the order of one second). Clearly, investigating either of these
scenarios requires time--dependent, multidimensional calculations with
detailed microphysics
\citep{rjs96,kl98,rj99,lee01,rrr02,rrrd03,rsw04}. The energy released
by accretion is then transformed into a relativistic outflow which
produces the gamma ray burst \citep[see][for
reviews]{m02,zm04,p04}. We note that either scenario would produce a
distinct gravitational wave signal, which could in principle be
detected by interferometric systems such as LIGO in the future, thus
establishing the nature of the progenitor system without a doubt.

The regime in which the gas lies is unlike any other commonly
encountered in astrophysics. The high densities and temperatures lead
to photodisintegration of the nuclei and the establishment of nuclear
statistical equilibrium (NSE). Furthermore, neutronization becomes
important in the innermost regions of the flow and weak interactions
determine the composition, with the electron fraction falling
substantially below 1/2, and the gas correspondingly becoming
neutron--rich. If this composition is somehow frozen and transported
out of the gas and into an outflow, interesting nucleosynthesis of
heavy elements could occur \citep{qw96,pwh03,pth04}. Realistic physics
input of this kind allow us to obtain more reliable estimates of the
actual energy released from the disk, and potentially available to
power a GRB. An added complication, which affects the composition
\citep{bel03} is that even if the main cooling mechanism is neutrino
emission, these are not entirely free to leave the system, as
scattering (mainly off free nucleons) is important enough to suppress
the emission in the dense inner disk. We find that in fact the
opaqueness of the material may lead to convection through the
establishment of a composition gradient that does not satisfy the
classical requirements for stability. To our knowledge, this is the
first time that this has been addressed in this context.

In previous work we initially studied the merger process for black
hole--neutron star binaries in three dimensions, paying particular
attention to the structure of the accretion disks that formed as a
result of the tidal disruption \citep{lee01}. Follow--up work used the
results of these simulations, mapped to two dimensions in azimuthal
symmetry, as initial conditions with simple (ideal gas) input physics,
and no realistic cooling included \citep{lrr02}. A first approximation
at a realistic equation of state and the effects of neutrino opacities
was reported more recently \citep{lrrp04}.

In this paper we improve upon our earlier results in three important
ways, that make them more realistic, and particularly relevant to the
study of GRBs. First, we use a much more detailed equation of state,
appropriate for the actual physical conditions found in post--merger
accretion disks. This includes a relativistic electron gas of
arbitrary degeneracy, radiation pressure and an ideal gas of free
nucleons and $\alpha$ particles. The composition is determined
self--consistently by considering weak interactions in two different
regimes: neutrino--opaque and neutrino--transparent. Second, we
include neutrino emission as the main source of cooling, by
considering the relevant reaction rates (electron and positron capture
onto free nucleons, bremsstrahlung, pair annihilation and plasmon
decays), taken from tables and fitting formulae valid over wide ranges
of temperature and density whenever possible. Third, we compute an
approximate optical depth for the fluid to neutrinos, using scattering
off free nucleons and $\alpha$ particles as a source of opacity. The
emission rates and pressure due to neutrinos are then suppressed and
enhanced, respectively, by an appropriate factor.

We begin this paper by a presentation of the physical conditions
likely to occur in post--merger accretion disks in \S\ref{pmdisks},
immediately following the coalescence of the two compact objects. Our
results follow in \S\ref{results}, and a discussion of these, and the
implications they have for GRBs is presented in \S\ref{ccl}.

\section{Post--merger accretion disks}\label{pmdisks}

The accurate study of the dynamical evolution of the accretion disks
requires knowledge of the physical conditions within them, and the use
of an appropriate equation of state. Below we give a general overview
of the physical conditions in the disk, followed by a detailed
presentation of the equation of state and the effects of neutrinos.

\subsection{Physical conditions}\label{cond}

The accretion structures that are formed as a result of the merger of
two neutron stars or the tidal
disruption of a neutron star by the black hole become azimuthally
symmetric fairly quickly, within a few tens of milliseconds
\citep{rjs96,lee01}. These disks typically contain a few tenths of a solar
mass, and are small, with the bulk of the mass being contained within
200~km of the black hole (which harbors about 3--5 solar masses). The
disks are dense ($10^{9}\leq\rho \mbox{[g~cm$^{-3}$]}\leq 10^{12}$),
with high internal energies ($10^{10}\leq T \mbox{[K]}\leq10^{11}$)
\citep[see e.g.,][]{rj99,retal99,lrr02}. In fact, the temperature is
high enough that nuclei become photodisintegrated and there is a
mixture of $\alpha$ particles, free neutrons and protons, electrons
and positrons. The timescale for $\beta$--equilibrium is given by
$t_{\beta} \approx (\sigma_{Ne} n_{\rm N} c)^{-1}$, where
$\sigma_{Ne}$ is the cross section for $e^{\pm}$ capture onto free
nucleons and $n_{\rm N}$ is the number density of free nucleons
\citep[see e.g.,][]{st83}. Under these conditions, $t_\beta \approx 2
\times 10^{-4}$~s, which is much shorter than the accretion timescale
$t_{acc}$ of a fluid element, so that weak interactions determine the
composition. Photons are trapped and are advected with the flow. For
neutrinos the situation is more complicated, since the optical depth
is of order unity. In the outer regions of the disk, they escape
freely, whereas for densities larger than $\approx
10^{11}$~g~cm$^{-3}$ they undergo diffusion on a timescale $t_{\rm
dif,\nu}\approx 30$~ms.

\subsection{The equation of state and the composition of the fluid.}
\label{eoscomp}

For simplicity of presentation, we will first assume that all nucleons
are free (no $\alpha$ particles are present), and include the
necessary corrections subsequently.

Under the conditions described above, the temperature is high enough
for electron--positron pair creation. The number of pairs thus
produced is very sensitive to the degeneracy of the electrons. In fact
the number density of pairs is suppressed exponentially with
increasing degeneracy, because of Fermi blocking. In previous work
\citep{pwf99,npk01,km02,dpn02,lrrp04} the assumption of full
degeneracy has been made for simplicity (note also that in some cases
the presence of electron--positron pairs has been assumed while at the
same time retaining a degeneracy pressure term, which is
inconsistent). Here we take a different approach, using an exact
expression for the pressure as a function of the temperature and the
chemical potential, valid for arbitrary degeneracy in the limit of
relativistic electrons (i.e., $\rho \geq 10^{6}$~g~cm$^{-3}$), due to
\citet{blinnikov96}, namely:

\begin{equation}
P_{e}=\frac{1}{12 \pi^{2} (\hbar c)^{3}}\left[
\eta_{e}^{4}+2\pi^{2}\eta_{e}^{2}(kT)^{2}+\frac{7}{15}\pi^{4}(kT)^{4}\right]. 
\label{pe}
\end{equation}
The number densities of electrons and positrons are related by:
\begin{equation}
\frac{\rho Y_{e}}{m_{u}}=n_{-}-n_{+}=\frac{1}{3\pi^{2}(\hbar
c)^{3}}[\eta_{e}^{3}+\eta_{e} \pi^{2}(kT)^{2}] \label{ye}
\end{equation}
where $Y_{e}$ is the electron fraction. The chemical potential of
species $i$ is denoted by $\eta_{i}$ throughout. This expression
reduces to the well--known limits when the temperature is low ($kT \ll
\eta_{e}$, which gives $P\propto \rho^{4/3}$, appropriate for a cold
relativistic Fermi gas) and when it is high ($kT \gg \eta_{e}$, which
gives $P\propto T^{4}$, when the pressure comes from relativistic
electron--positron pairs). The full equation of state then reads:
\begin{equation}
P=P_{\rm rad}+P_{\rm gas}+P_{e}+P_{\nu}, \label{eos}
\end{equation}
where
\begin{equation}
P_{\rm rad}=\frac{aT^{4}}{3},
\end{equation}
\begin{equation}
P_{\rm gas}=\frac{\rho k T}{m_{u}},
\end{equation}
and $P_{\nu}$ is the pressure due to neutrinos (discussed below).
Here $a$ is the radiation constant, $k$ is Boltzmann's constant and
$m_{u}=1.667 \times 10^{-24}$~g is the atomic mass unit. Since the
presence of pairs is automatically taken into account in the
expression for $P_{e}$, there is no alteration to the numerical factor
$1/3$ in the expression for $P_{\rm rad}$. For the conditions encountered
in the accretion disks presented below, gas pressure dominates at the
80\% level over the other terms.  Regarding the photons, since the
temperature is $T \approx 5$~MeV, the peak in the blackbody spectrum
is at $\approx 14$~MeV. On the other hand, the plasma frequency,
$\omega_{p}$, corresponds to $T_{p}\approx 0.5$~MeV. Thus the standard
expression for radiation pressure may be considered accurate, and
plasma effects negligible as far as the photons are concerned.

The computation of the electron fraction and the chemical potential of
electrons follows from the assumption of $\beta$--equilibrium between
neutrons, protons and electrons, and the condition of charge
neutrality. As noted by \citep{bel03}, a distinction needs to be made
to determine the equilibrium composition depending on the optical
depth of the material (the determination of the opacities is presented
in \S~\ref{nuopacities}). If it is transparent to its own neutrino
emission, we fix equilibrium by equating the capture rates of
electrons and positrons onto protons and neutrons respectively. For
mild degeneracy, as is the case here, this leads to the following
expression for the electron fraction as a function of the temperature
and the electron chemical potential \citep{bel03}:
\begin{equation}
Y_{e}=\frac{1}{2}+0.487 \left(\frac{Q/2-\eta_{e}}{kT} \right), \label{yethin}
\end{equation}
where $Q=(m_{n}-m_{p})c^{2}\simeq 1.29$~MeV.
If, however, the material is opaque, and the neutrinos are allowed to
diffuse out on a timescale shorter than the accretion timescale, then
we may write
\begin{equation}
\eta_{e}+\eta_{p}=\eta_{n}, \label{chemeq}
\end{equation}
for the equilibrium composition\footnote{This expression neglects 
the neutrino chemical potential, $\eta_{\nu}$, and is 
strictly valid only when there are equal numbers of neutrinos and 
anti--neutrinos. Otherwise the electron pressure, bulk viscosity and 
Joule--like heating due to the induced lepton current will be 
affected \citep{bml81,sbhf04}. We neglect all these corrections 
in the present treatment.}. Since the nucleons are not
degenerate, we may use Maxwell--Boltzmann statistics to describe their
distribution function, and obtain
\begin{equation}
\frac{n_{p}}{n_{n}}=\exp[(Q-\eta_{e})/kT] \label{pnfrac} 
\end{equation}
for the ratio of proton to neutron number densities. 
Further, with $Y_{e}=n_{p}/(n_{p}+n_{n})$ we arrive at
\begin{equation}
\frac{1-Y_{e}}{Y_{e}}=\exp[(\eta_{e}-Q)/kT]. \label{yethick}
\end{equation}

We now have the three equations~(\ref{ye}), (\ref{eos}) and
(\ref{yethin}) or (\ref{yethick}) for the three functions $T, Y_{e},
\eta_{e}$ and so the system is closed. To allow for a transition from
the optically thin to optically thick regime, in fact we solve these
in a combined form, weighted by factors
$f(\tau_{\nu})=\exp[-\tau_{\nu}]$ or
$g(\tau_{\nu})=(1-\exp[-\tau_{\nu}])$. As a practical matter, the
internal energy per unit mass, $u$, is used instead of the pressure to
solve equation (\ref{eos}), since its variation in time is what is
determined in the code, using the First Law of Thermodynamics. We
finally include the effects of incomplete photodisintegration of
$\alpha$ particles by using nuclear statistical equilibrium for the
three species $(n,p,\alpha)$ to fix the mass fraction of free nucleons
as \citep{qw96}:
\begin{equation}
X_{\rm nuc}=22.4 \, \left( \frac{T}{[10^{10}\mbox{K}]}\right)^{9/8} \,
\left(\frac{\rho}{10^{10}\mbox{g~cm$^{-3}$}}\right)^{-3/4} \, \exp(-8.2
[10^{10}\mbox{K}]/T).\label{nuc}
\end{equation}
Whenever this expression results in $X_{\rm nuc}>1$ we set
$X_{\rm nuc}=1$. The corresponding alterations in the previous derivation
are simple, and lead to the full set of equations which we write for
each gas element and at each time. Note that these do not allow for an
explicit solution, so an iterative scheme is used in each
instance. The internal energy per unit mass is
\begin{equation}
u=3\frac{P_{e}+P_{\rm rad}+P_{\nu}}{\rho}+\frac{1+3X_{\rm
nuc}}{4}\frac{3kT}{2m_{u}},
\label{thermal}
\end{equation}
$\beta$--equilibrium gives
\begin{equation}
Y_{e}=(1-X_{\rm nuc})/2 + X_{\rm nuc} \left( \left[
\frac{1}{2}+0.487\left\{\frac{Q/2-\eta_{e}}{kT} \right\}\right]
f(\tau_{\nu})+[1+\exp(\{\eta_{e}-Q\})/kT)]g(\tau_{\nu}) \right) ,
\label{betaeq}
\end{equation}
and charge neutrality implies
\begin{equation}
\frac{\rho Y_{e}}{m_{u}}=\frac{1}{3\pi^{2}(\hbar
c)^{3}}[\eta_{e}^{3}+\eta_{e} \pi^{2}(kT)^{2}],
\label{neutrality}
\end{equation}
as given by equation~(\ref{ye}). The modifications to the equation for
$\beta$ equilibrium simply reflect the fact that if the fluid is
composed primarily of $\alpha$ particles, there will be one neutron
per proton, and hence $Y_{e}\rightarrow 1/2$.

\subsection{Neutrino emission and photodisintegration losses}
\label{nuemission}

Several processes contribute to the emission of neutrinos. First, we
take into account electron and positron capture onto free nucleons
using the tables of \citet{lmp01}. This is an improvement over
calculations done previously \citep{pwf99,npk01,km02,dpn02,lrrp04},
where assumptions concerning the degeneracy were made. The rates are
thus accurate over the entire disk, whether the degeneracy is
significant or not. The tables cover the following ranges: $1\leq
\log(\rho Y_{e} \mbox{[g~cm$^{-3}$]}) \leq 11$; $7\leq \log T
\mbox{[K]}\leq 11$. A bilinear interpolation in the $\log \rho
Y_{e}-\log T$ plane is performed using the table to obtain the cooling
rate for a given mass element. The result is then multiplied by the
mass fraction of free nucleons, $X_{\rm nuc}$, since we are not
considering capture of electrons and protons by $\alpha$
particles. Second, the annihilation of electron--positron pairs is a
source of thermal neutrinos, and the corresponding cooling rate is
computed using the fitting functions of \citet{itoh96}. These cover
the range $9\leq \log(\rho \mbox{[g~cm$^{-3}$]}) \leq 12$ in density
and $10\leq \log T \mbox{[K]}\leq 11$ in temperature. Finally,
nucleon--nucleon bremsstrahlung, $\dot{q}_{ff}$, and plasmon decay,
$\dot{q}_{\rm plasmon}$, are considered, with rates given by:
\begin{equation}
\dot{q}_{ff}=1.5\times 10^{33} \, T_{11}^{5.5} \, \rho_{13}^{2}\, 
\mbox{erg~s$^{-1}$~cm$^{-3}$}, \label{qff}
\end{equation}
\citep{hr98} and
\begin{equation}
\dot{q}_{\rm plasmon}=1.5\times 10^{32} \, T_{11}^{9} \,
\gamma_{p}^{6} \, \exp(-\gamma_{p}) (1+\gamma_{p})\left(
2+\frac{\gamma_{p}^{2}}{1+\gamma_{p}}\right)
\mbox{erg~s$^{-1}$~cm$^{-3}$},\label{qplasmon}
\end{equation}
where $\gamma_{p}=5.5 \times 10^{-2} \sqrt{(\pi^{2}
+3[\eta_{e}/kT]^{2})/3}$
\citep{rjs96}.  For the conditions encountered in the disk, capture by
nucleons completely dominates all other processes, and is the main
source of cooling.  Finally, the creation and disintegration of
$\alpha$ particles leads to a cooling term in the energy equation given by:
\begin{equation}
\dot{q}_{\rm phot}=6.8\times 10^{18} \,
\frac{dX_{\rm nuc}}{dt}~\mbox{erg~s$^{-1}$~cm$^{-3}$}.
\end{equation}

\subsection{Neutrino opacities}\label{nuopacities}

The material in the disk is dense enough that photons are completely
trapped, and advected with the flow. For neutrinos however, the
situation is more complicated, since the outer regions are transparent
to them, while the inner portions are opaque.

The main source of opacity is scattering off free nucleons and
$\alpha$ particles, with cross--sections given by \citep{ts75,st83}
\begin{equation}
\sigma_{\rm N}=\frac{1}{4}\sigma_{0}\left(
\frac{E_{\nu}}{m_{e}c^{2}}\right)^{2},
\end{equation}
and
\begin{equation}
\sigma_{\alpha}=\sigma_{0}\left(
\frac{E_{\nu}}{m_{e}c^{2}}\right)^{2}\left[ (4
\sin^{2}\theta_{w})\right]^{2},
\end{equation}
where $\sigma_{0}=1.76\times10^{-44}$~cm$^{2}$ and $\theta_{w}$ is the
Weinberg angle. Since capture processes dominate the neutrino
luminosity, we assume that the mean energy of the neutrinos is roughly
equal to the Fermi energy of the partially degenerate electrons
$E_{\nu} \approx \frac{5}{6}\eta_{e}=43 (Y_{e}\rho_{12})^{1/3}$~MeV,
where $\rho_{12}=\rho/10^{12}$g~cm$^{-3}$. So the mean free path is
$l_{\nu}=1/(n_{\rm N} \sigma_{\rm N} + n_{\alpha}\sigma_{\alpha})$,
$n_{\rm N}$ and $n_{\alpha}$ being the number densities of free
nucleons and $\alpha$ particles respectively. To compute an optical
depth, we take $\tau_{\nu}=H/l_{\nu}$, with $H$ a typical scale height
of the disk. Inspection of the disk shape (through the density
contours in the inner regions) and an estimation of the scale height
with $H\simeq \rho/\nabla \rho$ suggests that $H\simeq \kappa r$, with
$\kappa$ a constant of order unity, so that our final, simplified
expression for the optical depth reads $\tau_{\nu}=\kappa
r/l_{\nu}$. With the above expressions for the cross section as a
function of density and electron fraction, and defining
$r_{7}=r/10^{7}\mbox{cm}$ we arrive at
\begin{equation}
\tau_{\nu}=186.5 \, \kappa \, \rho_{12}^{5/3} \, Y_{e}^{2/3} \, r_{7}
\left[X_{\rm nuc}+3.31(1-X_{\rm nuc})/4 \right].
\label{eq:nudepth}
\end{equation}
The term which depends on $X_{\rm nuc}$ reflects the two--species
composition we have assumed. Its influence is limited, since when
$X_{\rm nuc}$ varies from 0 to 1, the optical depth is altered by a factor
1.2, all other values being equal. In practice, and for the
compositions found in the disks, this expression corresponds to having
an optical depth of unity at approximately $10^{11}$g~cm$^{-3}$, which
can be thought of as a neutrino--surface, where the last scattering
occurs and neutrinos leave the system. Accordingly, we modify the
expressions for the neutrino emission rates, suppressing it in the
opaque regions, and enhancing the pressure through
\begin{equation}
\left(\frac{du}{dt}\right)_{\nu}=\left(\frac{du}{dt}\right)_{0}
\exp(-\tau_{\nu}),
\end{equation}
and
\begin{equation}
P_{\nu}=\frac{7}{8}aT^{4}[1-\exp(-\tau_{\nu})],
\end{equation}
where $(du/dt)_{0}=\sum \dot{q}_{i}/ \rho$ is the unmodified energy
loss rate (in erg~g$^{-1}$~s$^{-1}$) calculated from the rates given
in \S~\ref{nuemission}.

The total neutrino luminosity (in erg~s$^{-1}$) is then computed
according to
\begin{equation}
L_{\nu}=\int \rho^{-1} \, \left[\dot{q}_{ff}+\dot{q}_{\rm
plasmon}+\dot{q}_{\rm pair}+ \dot{q}_{\rm cap} \right] \,
\exp(-\tau_{\nu}) \, dm. 
\label{eq:nulum}
\end{equation}

\subsection{Numerical method and initial conditions.}\label{method}

For the actual hydrodynamical evolution calculations, we use the same
code as in previous work \citep{lrr02}, and refer the reader to that
paper for the details. This is a two dimensional (cylindrical
coordinates $(r,z)$ in azimuthal symmetry) Smooth Particle
Hydrodynamics code \citep{monaghan92}, modified from our own 3D
version used for compact binary mergers \citep{lee01}. The accretion
disk sits in the potential well of a black hole of mass $M_{\rm BH}$,
and which produces a Newtonian potential $\Phi=-GM_{\rm
BH}/r$. Accretion is modeled by an absorbing boundary at the
Schwarzschild radius $r_{Sch}=2GM_{\rm BH}/c^{2}$, and the mass of the
hole is updated continuously. The transport of angular momentum is
modeled with an $\alpha$ viscosity prescription, including {\em all}
components of the viscous stress tensor (not only $t_{r\phi}$). The
self--gravity of the disk is neglected.

The main modifications from the previous version are in the
implementation of a new equation of state \S\ref{eoscomp}, neutrino
emission, \S\ref{nuemission} and the treatment of neutrino optical
depths, \S\ref{nuopacities}.

As done previously \citep{lrr02,lrrp04}, the initial conditions are
taken from the final configuration of 3D calculations of black
hole--neutron star mergers, after the accretion disk that forms
through tidal disruption of the neutron star has become fairly
symmetric with respect to the azimuthal coordinate. We show in
Table~\ref{table:ICs} the parameters used in each of the dynamical
runs included in this paper.

\section{Results}\label{results}

As the dynamical evolution calculations begins, the disks experience
an initial transient, which is essentially numerical in origin, and is
due to the fact that the configuration is not in strict hydrostatic
equilibrium (recall that it is obtained from azimuthally averaging the
results of 3D calculations). This transient, and its effects, are
negligible, and it is essentially over in a hydrodynamical timescale
($\approx 2$~ms). After that, the disk proceeds to evolve on a much
longer timescale, determined by accretion onto the central black
hole. We first present the details of the spatial structure of the
disk due to fundamental physical effects, and then proceed to show the
temporal evolution and associated transitions.

\subsection{Disk structure}\label{structure}

\begin{figure}
\plotone{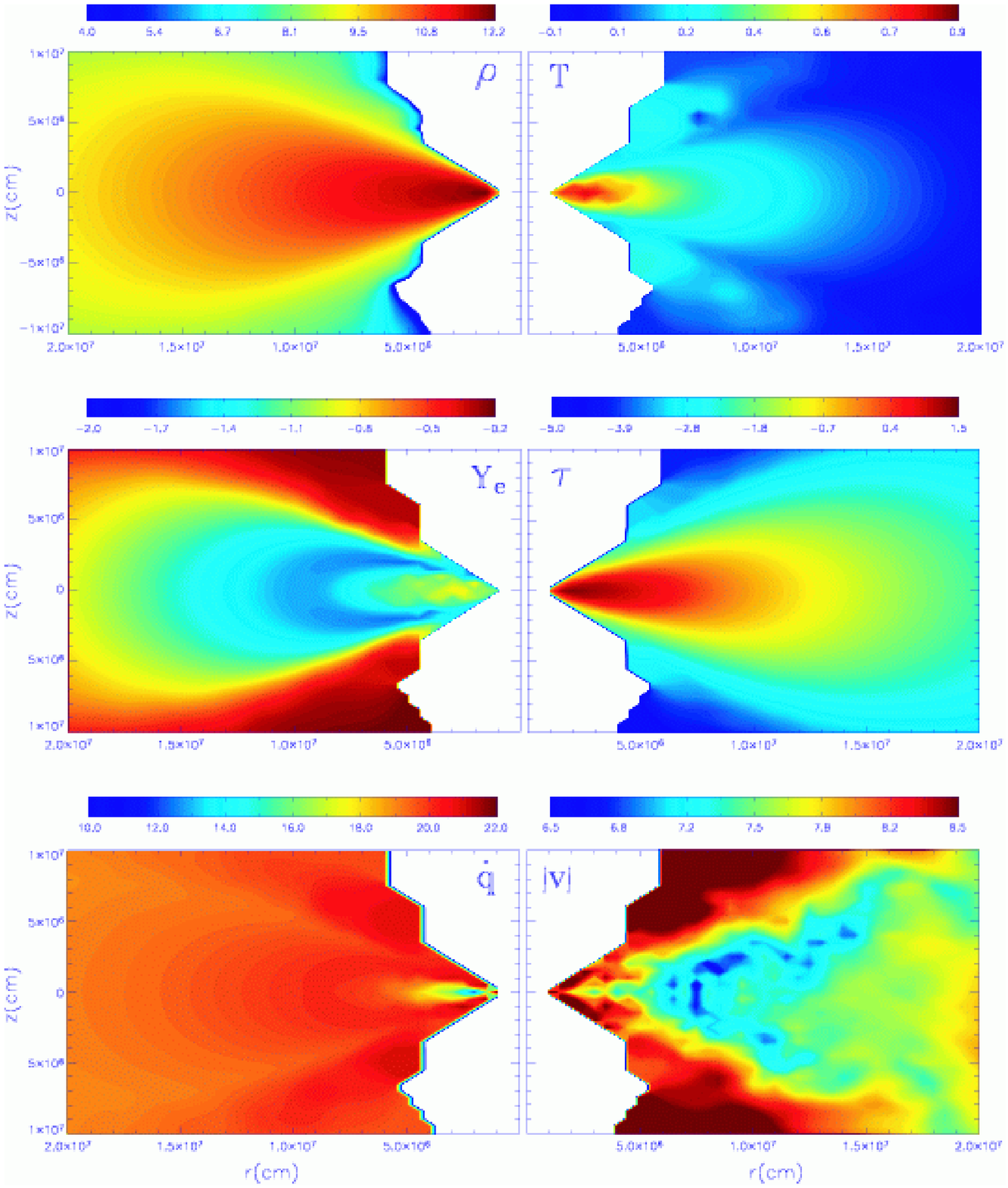}
\caption{Color coded logarithmic contours of density (g~cm$^{-3}$),
temperature (MeV), electron fraction, optical depth, cooling
(erg~g$^{-1}$~s$^{-1}$) and magnitude of velocity (cm~s$^{-1}$) for
run a2M at $t=100$~ms. The qualitative change in composition is
clearly seen in the contours of $Y_{e}$ when the material becomes
optically thick (note also the exponential suppression of cooling in
the contours of $\dot{q}$ in this regime). See text for details.}
\label{compa0.01t100}
\end{figure}

\begin{figure}
\plotone{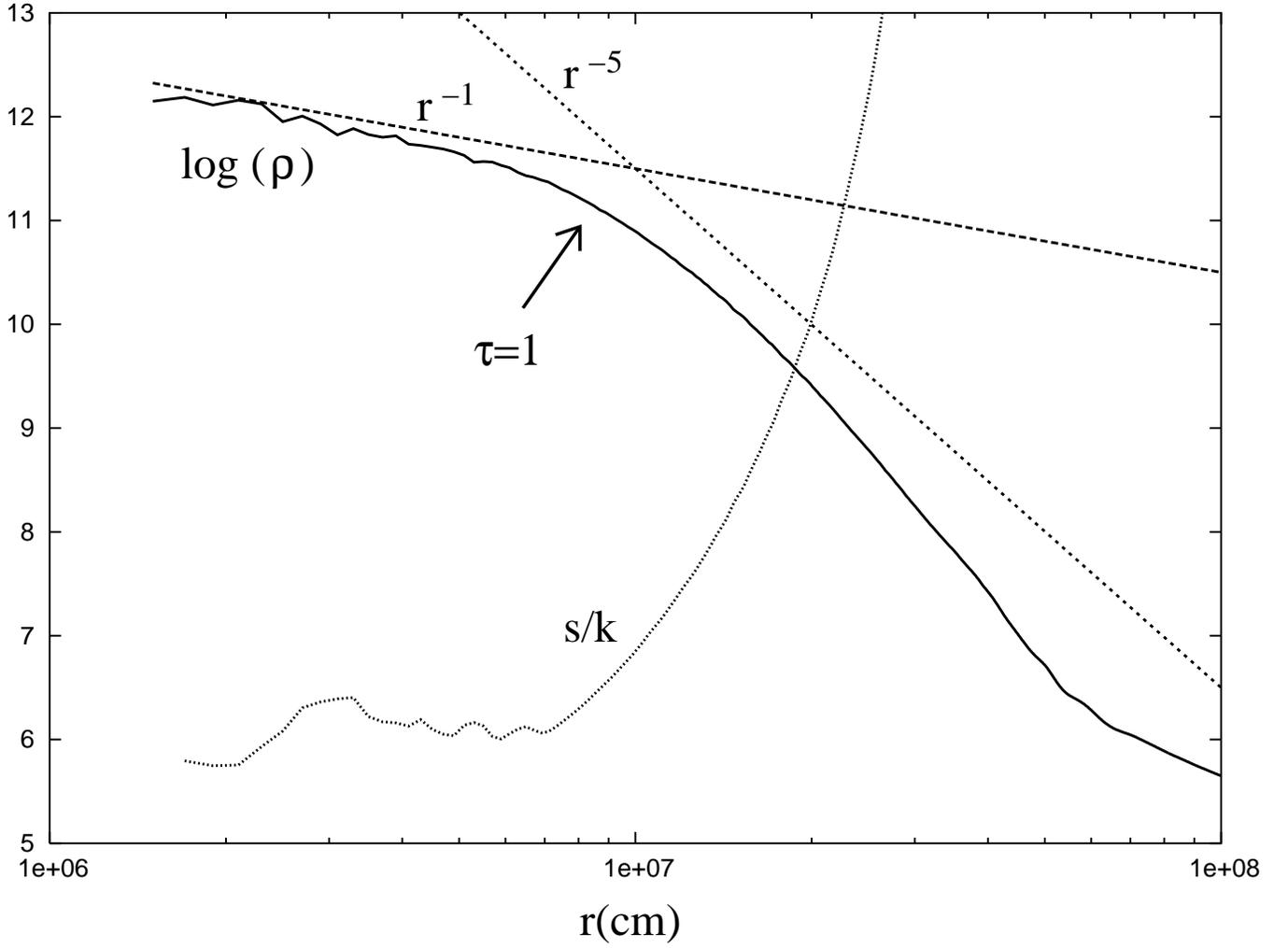}
\caption{Density and entropy per baryon, $s/k$, along the equatorial
plane, $z=0$, for run a2M at $t=100$~ms, as in
Figure~\ref{compa0.01t100}. Note the logarithimc scale in density and
the sharp transition and change in behavior from the optically thin to
optically thick regimes. Reference power laws for the density profile
are indicated. In the optically thick region, the entropy per baryon
is practically constant.}
\label{compa0.01t100r}
\end{figure}

The fundamental variable affecting the instantaneous spatial structure
of the disk is the optical depth to neutrinos, $\tau_{\nu}$, since it
determines whether the fluid cools efficiently or not. All quantities
show a radical change in behavior as the threshold $\tau_{\nu}=1$ is
passed. Figure~\ref{compa0.01t100} shows color--coded contours in a
meridional slice of the thermodynamical variables, while
Figure~\ref{compa0.01t100r} displays the run of density and entropy
per baryon along the equator, $z=0$, where the density and temperature
are highest (we will refer here mainly to run a2M, unless noted
otherwise). The density increases as one approaches the black hole,
varying as $\rho \propto r^{-5}$ in the outer regions, where
$\tau_{\nu} \ll 1$. The critical density for opaqueness is reached at
$r_{*}\simeq 10^{7}$~cm, and for $r<r_{*}$, $\rho \propto
r^{-1}$. Since the energy that would otherwise be lost via neutrino
emission remains in the disk if the cooling is suppressed by
scattering, the density does not rise as fast. In fact, one can note
this change also by inspecting the scale height $H\simeq P/ \nabla P$,
which scales as $H/r \propto r$ for $r>r_{*}$ and $H/r \propto$~const
for $r<r_{*}$ (and thus in terms of surface density the change is from
$\Sigma \propto r^{-4}$ to $\Sigma \propto$~const at $r_{*}$). The
opaque region of the disk remains inflated to a certain extent,
trapping the internal energy it holds and releasing it only on a
diffusion timescale. At densities $\rho \simeq 10^{12}$~g~cm$^{-3}$
the optical depth can reach $\tau_{\nu} \simeq 100$, and thus the
suppression of cooling is dramatic, as can be seen in
Figure~\ref{compa0.01t100}, where the contours of $\tau_{\nu}$ and
$\dot{q}$ are shown. The flattening of the entropy profile is related
to the change in composition in the optically thick region and the
occurrence of convection (see \S~\ref{nuconvection} below).

As the density rises in the inner regions of the disk, the electron
fraction $Y_{e}$ initially drops as neutronization becomes more
important, with the equilibrium composition being determined by the
equality of electron and positron captures onto free neutrons and
protons (see equation~(\ref{yethin}) and the discussion preceding
it). The lowest value is reached at $r \simeq r_{*}$, where $Y_{e}
\simeq 0.03$. Thereafter it rises again, reaching $Y_{e} \simeq 0.1 $
close to the horizon. Thus flows that are optically thin everywhere
will reach a higher degree of neutronization close to the black hole
than those which experience a transition to the opaque regime. The
numerical values for the electron fraction at the transition radius
and at the inner boundary are largely insensitive to $\alpha$, as long
as the transition does occur.

\begin{figure}
\plotone{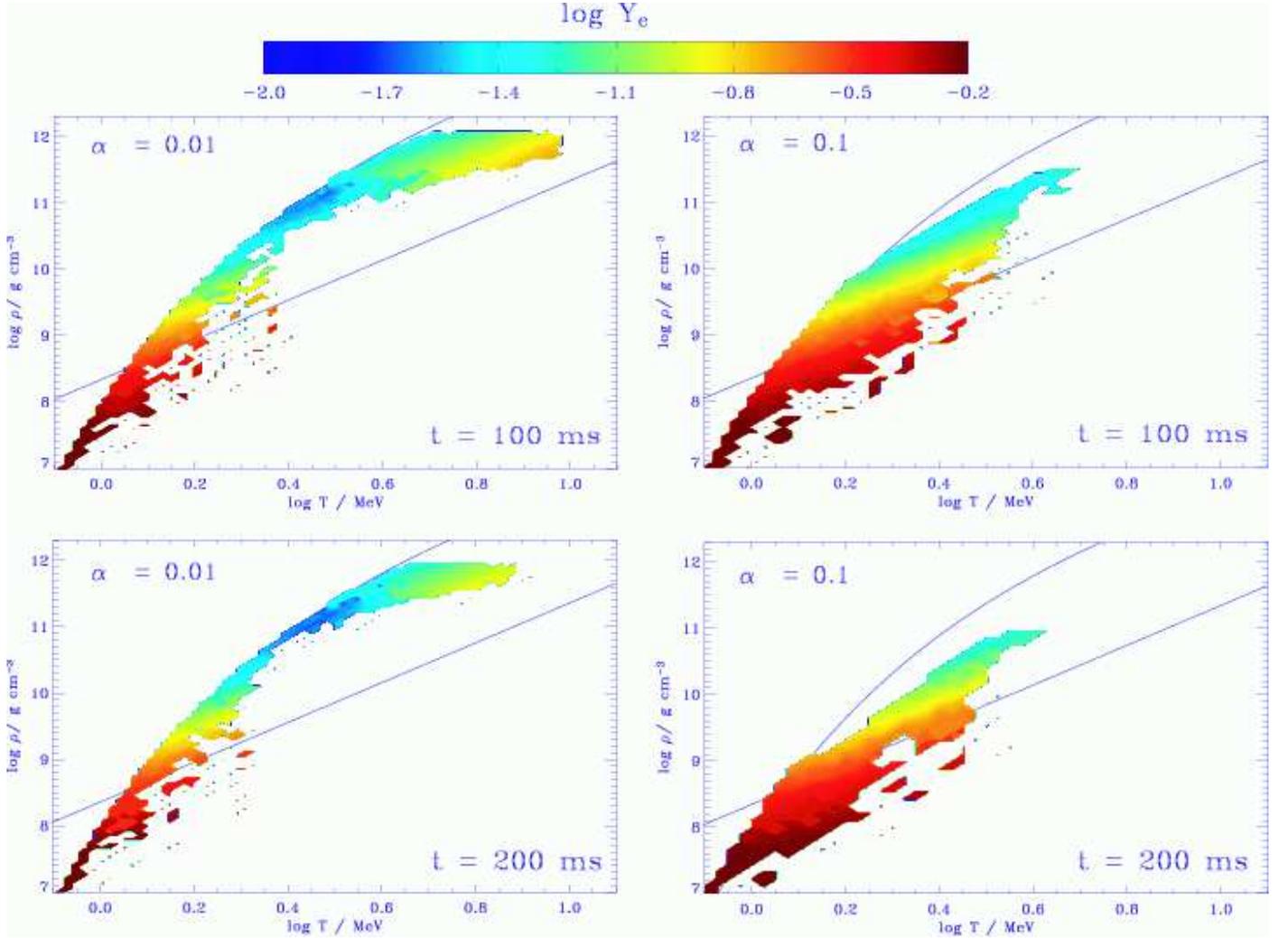}
\caption{Color coded electron fraction in the $\rho-T$ plane for runs
a2M (left column) and a1M (right column) at $t=100$~ms (top) and
$t=200$~ms (bottom). The solid curving line marks the transition in
composition from free nucleons (at high temperatures and low
densities) to $\alpha$ particles (at low temperatures and high
densities). The solid straight line marks the degeneracy threshold,
given by $kT=7.7\rho_{11}^{1/3}$~MeV. }
\label{rtye}
\end{figure}

The baryons in the disk are essentially in the form of free neutrons
and protons, except at very large radii and low densities ($r > 5
\times 10^{7}$~cm and $\rho < 5\times 10^{6} $g~cm$^{-3}$) where
$\alpha$ particles form. Figure~\ref{rtye} shows the region in the
density--temperature plane where the fluid lies, for runs a2M and a1M
at two different times (and color coded according to the electron
fraction). Most of the gas lies close to the line determining the
formation of Helium nuclei (see equation~\ref{nuc}), but does not
cross over to lower temperatures. The reason for this is that the
energy that would be released by the creation of one Helium nucleus
(28.3~MeV) would not leave the disk (recall that we consider neutrino
emission as the only source of cooling) and thus immediately lead to
the photodisintegration of another $\alpha$ particle. An equilibrium
is thus maintained in which the gas is close to Helium synthesis, but
this does not occur. In the opaque regime, the gas moves substantially
farther from the transition line to Helium, since it cannot cool
efficiently and, as mentioned before, the density does not rise as
quickly. Figure~\ref{rtye} also shows clearly why one must make use of
an equation of state which takes into account properly the effects of
arbitrary degeneracy of the electrons and positrons. The solid
straight line marks the degeneracy temperature as a function of
density, given by $kT=7.7 \rho_{11}^{1/3}$~MeV. The disk straddles
this line, with a degeneracy parameter $\eta_{e}/kT \simeq 2-4$ in the
inner regions, and $\eta_{e}/kT \approx 1$ at lower densities. Thus
making the approximation that the electrons in the flow are fully
degenerate is not accurate.

\begin{figure}
\plotone{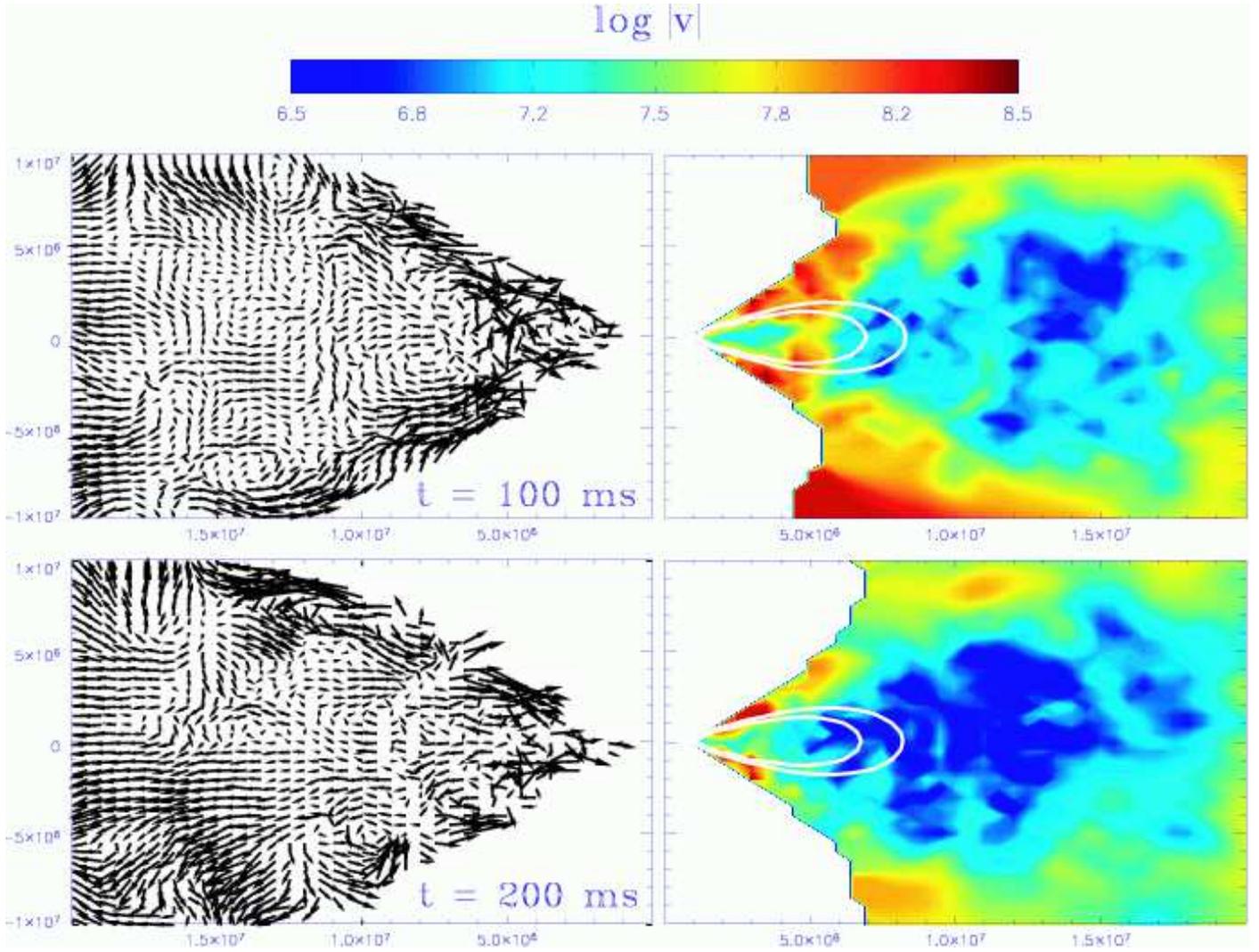}
\caption{Velocity field (left column) and color coded velocity
contours (right column) for run a2M at $t=100$~ms (top) and $t=200$~ms
(bottom), showing clearly the small scale circulations present in the
disk (the units are cm~s$^{-1}$). The solid white contours 
correspond to optical depths $\tau=3/2, 6$.}
\label{velcont}
\end{figure}

\begin{figure}
\plotone{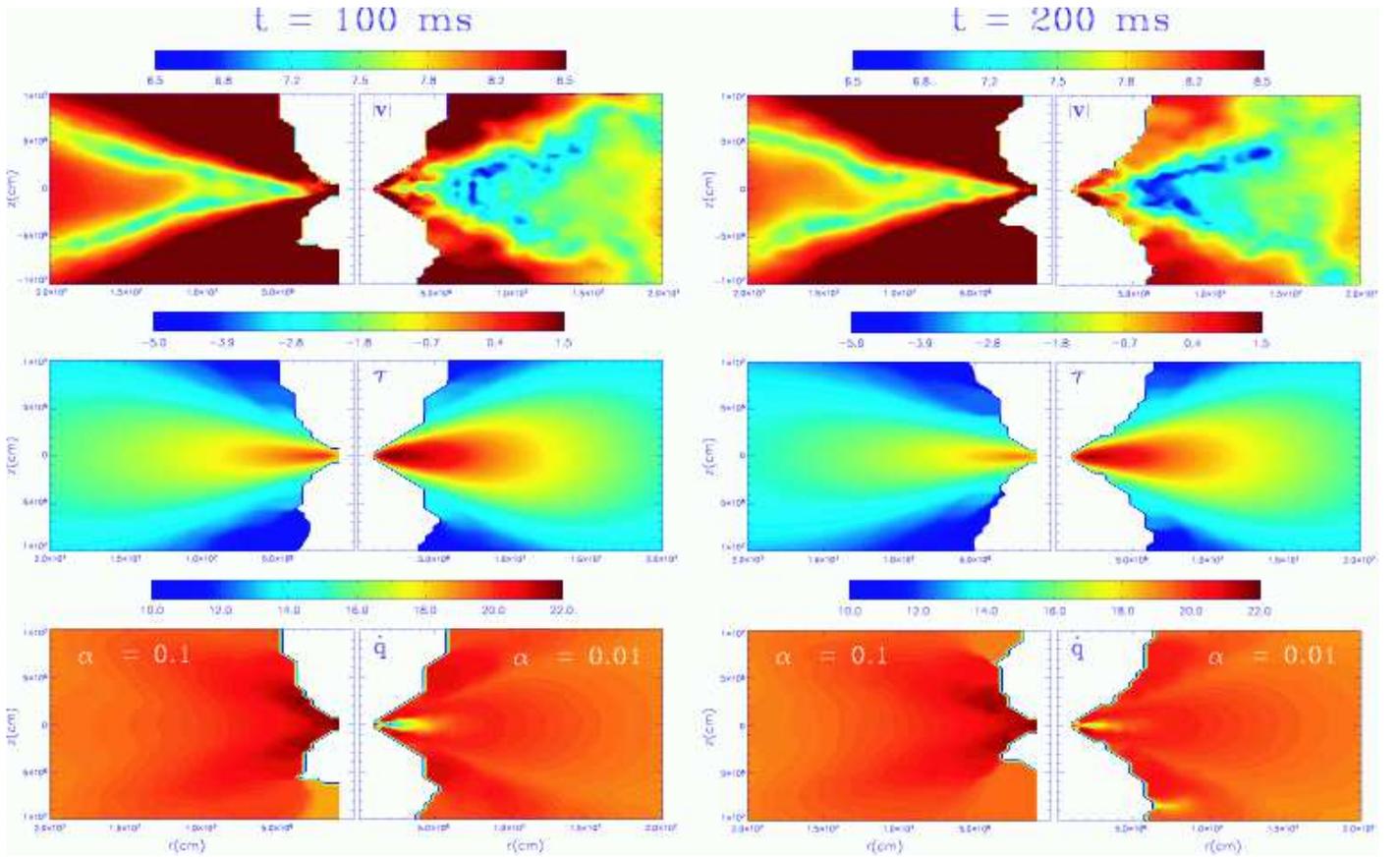}
\caption{Color coded logarithimic contours of magnitude of velocity
(top, in cm~s$^{-1}$), optical depth (middle) and cooling (bottom, in
erg~g$^{-1}$~s$^{-1}$ ) for runs a1M (left half of each panel) and a2M
(right half of each panel), at $t=100$~ms (left column) and $t=200$~ms
(right column). Note the strong thinning of the disk in the case with
high viscosity, $\alpha=0.1$, visible in the changing contours of
optical depth. }
\label{compa0.01a00.1t100t200}
\end{figure}

Modeling the accretion flow in two dimensions $(r,z)$, without the
assumption of equatorial symmetry allows one to solve clearly for the
vertical motions in the disk, something which is not possible when
considering vertically--integrated flows. It was pointed out by
\citet{urpin84} that a standard $\alpha$ viscosity could lead to
meridional flows in which $v_{r}$ changed sign as a function of height
above the mid plane, leading to inflows as well as
outflows. \citet{kita95}, \citet{kk00} and \citet{rg02} later
considered a similar situation, and found that for a range of values
in $\alpha$, the gas flowed inward along the surface of the disk, and
outward in the equatorial regions. In previous work \citep{lrr02}, we
found this solution in disk flow, with large scale circulations
exhibiting inflows and outflows for high viscosities ($\alpha \simeq
0.1$), and small--scale eddies at lower values ($\alpha \leq
0.01$). How vertical motions affect the stability of accretion disks
and may lead to the transport of angular momentum is a question that
has become of relevance in this context \citep{au04}. Here we show in
Figure~\ref{velcont} the velocity field and magnitude of meridional
velocity for run a2M. The small--scale eddies are clearly visible,
with their strength usually diminishing as the disk is drained of
matter and the density drops.  The effect of a different value of
$\alpha$ can be seen in the top row of
Figure~\ref{compa0.01a00.1t100t200}, where the comparison between
$\alpha=0.1$ and $\alpha=0.01$ is made. The large, coherent lines of
flow aimed directly at the origin for $\alpha=0.1$ correspond to
inflow, while the lighter shades along the equator at larger radii
show outflowing gas.

The instantaneous structure of the disk concerning the density,
temperature and composition profiles is largely independent of the
viscosity, as long as there is enough mass to produce the optically
thin/optically thick transition. Even in the optically thin regime,
the structure cannot be determined analytically following the standard
arguments applied to thin, cool disks of the Shakura--Sunyaev
type. The reason for this is the following: a central assumption in
the standard solution is that the disk is cool, i.e., $kT/m_{p} \ll
GM_{\rm BH}/r$. This comes from the requirement that all the energy dissipated
by viscosity, $\dot{q}_{\alpha}$ be radiated away efficiently, and
produce the observed flux, $F$. In our case, the disk is most
certainly not cool, as it originates from the tidal disruption of a
neutron star by a black hole (or possibly the coalescence of two
neutron stars, and subsequent collapse of the central mass to a black
hole). The gas that constitutes the disk is dynamically hot because it
was in hydrostatic equilibrium in a self--gravitating configuration,
where $U \simeq -W$ and has not been able to release this internal
energy (the merger process itself may lead to additional heating). A
further deviation from the standard solution is that the pressure
support in the disk leads to a rotation curve that is slightly
sub--Keplerian.  The dissipation by viscosity is in fact smaller than
the cooling rate over much of the disk, and the released luminosity
comes from a combination of viscous dissipation and the store of
internal energy given to the disk at its conception.

\subsection{Lepton--driven convection}\label{nuconvection}

The classical requirement for convective instability in the presence
of entropy and composition gradients, as well as rotation, is the
Solberg--H\/{o}iland criterion \citep{t78}:
\begin{equation}
N^{2}+\omega_{r}^{2} < 0, \label{SH}
\end{equation}
where $\omega_{r}^{2}=4 \Omega^{2} + r d\Omega^{2}/dr$ is the radial
epicyclic frequency ($\Omega$ being the angular velocity), and
\begin{equation}
N^{2}=\frac{g}{\gamma} \left[ \frac{1}{P} \left( \frac{\partial
P}{\partial s}\right)_{Y_{e}} \frac{ds}{dr} + \frac{1}{P} \left(
\frac{\partial P}{\partial Y_{e}} \right)_{s} \frac{dY_{e}}{dr}
\right] \label{BV}
\end{equation}
is the Brunt--V\"{a}is\"{a}la frequency \citep[see
][]{lm81,tqb05}. The adiabatic index $\gamma=d\ln P / d\ln \rho
\approx 5/3$ in our case, since $P_{gas}$ is the most important
contribution to the total pressure. Thus a region may be convectively
unstable because of a composition gradient, or an entropy gradient, or
both. Strictly speaking, here one should consider the total lepton
fraction $Y_{l}$. We do not consider the transport of neutrinos in
detail, as already mentioned (see \S~\ref{nuopacities}), and do not
calculate explicitly the contribution of neutrinos to this
$Y_{l}$. For what follows, we will thus simply take $Y_{e}$ to
represent the behavior of the full $Y_{l}$. The origin of convection
in the lepton inversion zone can be understood as follows
\citep{e79}. Consider a fluid element in the lepton inversion zone
that is displaced in the outward direction and then comes to pressure
equilibrium with its surroundings. The displaced element, which is
lepton rich relative to its new surroundings, attains the ambient
pressure at a lower density than the surrounding fluid, because the
pressure depends directly on the lepton number, and thus tends to
drift outwards. By the same token, an inwardly displaced fluid element
in the lepton inversion zone is depleted in leptons relative to its
new surroundings and thus tends to sink
\footnote{This is similar to the thermosolutal convection which would
occur in a normal star if there were an inwardly decreasing molecular
weight gradient or composition inversion \citep{spiegel72}.}.  When
temperature gradients are allowed for, the displaced fluids, in
general, have temperatures which differ from those of their
surroundings. These variations tend to promote stability or
instability depending on whether the existing temperature gradient is
less than or greater than the adiabatic gradient, respectively.

In the present scenario then, the transition to the optically thick
regime leads to instability, because initially $ds/dr < 0$, and also
$dY_{e}/dr <0$ (see also Figure~\ref{compa0.01t100r}). The entropy
profile is then flattened by efficient convective mixing, and the
lepton gradient inversion is due to the different way in which the
composition is determined through weak interactions once the neutrinos
become trapped. In a dynamical situation such as the one treated here,
convection tends to erase the gradients which give rise to
it. Accordingly, the sum of the corresponding terms in
equation~(\ref{SH}) tends to zero once the simulation has progressed
and convection has become established in the inner disk (note that in
our case the rotation curve there is sub--Keplerian, with $\Omega
\propto r^{-8/5}$, so that $\Omega$ and $\omega_{r}$ are not
equal). This is analogous to what occurs in proto neutron stars
following core collapse \citep{e79,bl86,td93} and has actually been
confirmed in numerical simulations of such systems \citep{jm96}, where
the convection leads to strong mixing. For a neutrino--driven
convective luminosity $L_{\rm con}$ at radius $R$ and density $\rho$,
the convective velocity may be estimated by standard mixing length
theory as \citep[see e.g.,][]{td93}
\begin{equation}
v_{\rm con} \sim 3.3 \times 10^{8} \left({L_{\rm con} \over
10^{52}\;{\rm erg\;s^{-1}}}\right)^{1/3} \left({\rho \over 10^{12}
{\rm g\;cm^{-3}}}\right)^{-1/3} \left({R \over 20\;{\rm
km}}\right)^{-2/3}\;\;{\rm cm\;s^{-1}},
\end{equation}
where we have used the fact that gas pressure dominates in the
fluid. The assumption of spherical symmetry implicit in this
expression is not strictly met in our case, but it may provide us
nevertheless with a useful guide. The overturn time of a convective
cell is then given by $t_{\rm con} \sim l_p/v_{\rm con}\sim 10$ ms,
where $l_p \sim 20$~km is the mixing length, usually set to a pressure
scale height.  In our calculations, we see that the magnitude of the
meridional velocity $|v|=\sqrt{v_{r}^{2}+v_{z}^{2}}$ decreases as $r$
decreases, then rises again as the opaque region is reached, reaching
$|v|\simeq 10^{8}$~cm~s$^{-1}$, at $r \sim 20$~km. The associated
turnover times are thus $t_{\rm con} \simeq l_p/|v| \simeq 20$~ms, in
good agreement with the estimate made above.

\begin{figure} 
\plotone{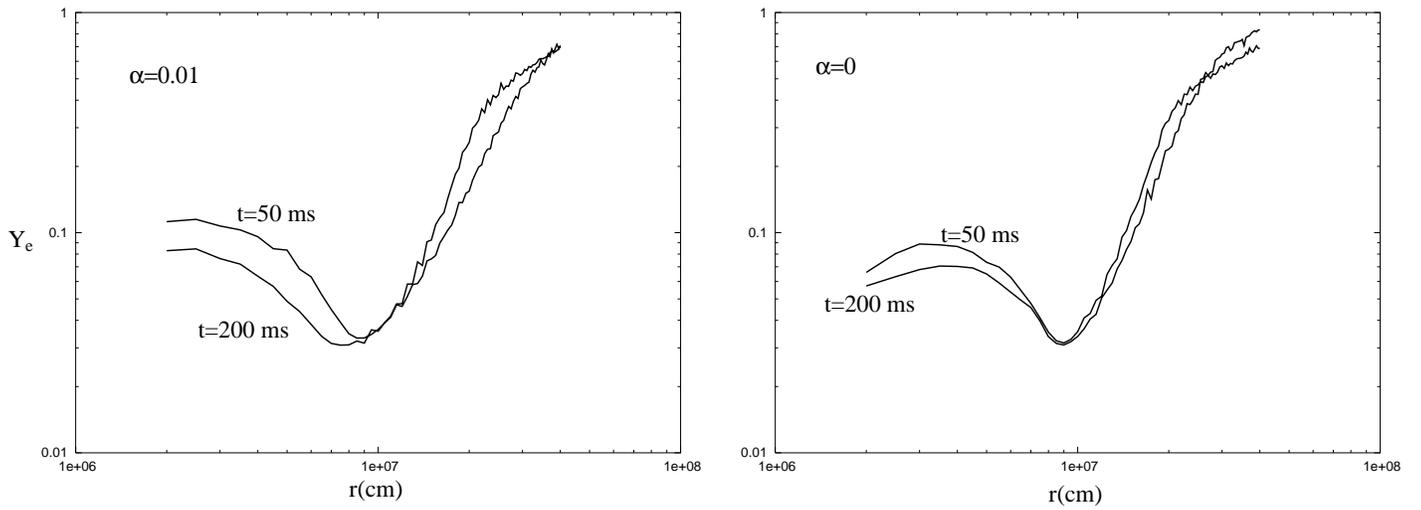}
\caption{Averaged profiles of the electron fraction $Y_{e}(r)$ for runs
a2M (left panel) at $t=50$~ms and $t=200$~ms and run aIM (right panel)
at $t=50$~ms and $t=200$~ms. In the inviscid calculation convection is
able to suppress the composition gradient in a few characteristic
turnover times, while the continuous radial motions induced by a
finite value of $\alpha$ prevent this from happening in the run with
viscosity.}
\label{convection}
\end{figure}

To isolate this effect from that due to viscosity (which generates the
meridional circulations mentioned previously) we have performed a
simulation in which $\alpha=0$ (run aIM in
Table~\ref{table:ICs}). This calculation shows essentially no
accretion onto the black hole except for a small amount of gas
transferred at early times (because the initial condition is not in
strict equilibrium). The result concerning the profile of $Y_{e}$ as a
function of radius is as we have described above, i.e., increasing
neutronization as $r$ decreases, until the opaque region is reached,
followed by an increase in $Y_{e}$ as the black hole boundary is
approached. The revealing difference lies in the time evolution of
this profile. As mentioned above, convection generates motions which
tend to eliminate the composition gradient which drives it. In the
presence of viscosity, matter is continuously transported radially,
and the gradient is not erased entirely. Height integrated profiles of
$Y_{e}(r)$ at 50~ms and 200~ms show a similar behavior at small radii.
When viscosity is removed, the initial composition gradient is
gradually softened until it disappears in the innermost regions (see
Figure~\ref{convection}). The disk then has a nearly constant electron
fraction along the equator for $r<r_{*}$, with a sharp transition
region leading to an increase with radius in the transparent regime.
This may resemble the convection dominated accretion solution found
analytically by \citet{qg00} and numerically by several groups
\citep{spb99,nia00}, where convection transports angular momentum
inward, energy outward, and gives a radial profile in density $\propto
r^{-1/2}$. We find a different power law, with $\rho \propto
r^{-1}$. There are several factors which may account for this
difference: the disk vs. spherical geometry; the initial condition
with a large amount of internal energy; and the limited, but present
cooling rate on the boundaries of the flow.

\subsection{Neutrino diffusion effects}

Since the neutrinos are diffusing out of the fluid (the mean free path
is small compared to the size of the system in the optically thick
regime), one would expect a corresponding viscosity through
$\zeta_{\nu}^{\rm vis}$ \citep[see, e.g.][]{bml81}. This needs to be
compared to the corresponding viscosity generated by the assumed
$\alpha$ prescription for consistency. We may assume $\zeta_{\nu}^{\rm
vis} = (1/3) U_{\nu} l_{\nu}/ c$, where $U_{\nu} \approx a T^{4}$ is
the energy density associated with neutrinos, and $l_{\nu}$ is their
mean free path (see \S~\ref{nuopacities}). Thus
\begin{equation} 
\zeta_{\nu}^{\rm vis}\approx 2
\times 10^{20} \mbox{g~cm$^{-1}$~s$^{-1}$}, \label{nuvisc}
\end{equation} 
for conditions near the equatorial plane, $z=0$. This value will
increase in the outer regions, since the mean free path becomes larger
as the density drops. At the neutrino--surface, where
$\tau_{\nu}\approx 1$, the mean free path is $l_{\nu}\approx
10^{7}$~cm, and so an averaged value of the viscosity over the entire
neutrino--opaque region will result in an increase of about one order
of magnitude over the estimate given in equation~(\ref{nuvisc}) (note
however, that strictly speaking the diffusion approximation is no
longer valid in the outer regions, so this must be interpreted with
care). For the $\alpha$ prescription, $\zeta_{\alpha}=\rho \alpha
c_{s}^{2}/\Omega$. The rotation curve is not too far from Keplerian,
and scaling this expression to typical values we find
\begin{equation} \zeta_{\alpha}=1.3 \times 10^{22} \,
c_{s,9}^{2} \, \alpha_{-2} \, r_{7}^{3/2} \, M_{4}^{-1/2} \, \rho_{10}
\, \mbox{g~cm$^{-1}$~s$^{-1}$}.  
\end{equation} 
The effects on angular momentum transport are thus a full two orders
of magnitude below those arising from our viscosity prescription (for
$\alpha=10^{-2}$). To put it another way, a lower limit for the
viscosity under these conditions (and in the optically thick portion
of the disk) would be $\alpha \simeq 10^{-4}$. An alternative analysis
would be to consider the corresponding timescales induced by this
viscosity. Since $t_{\rm vis}\propto R (\rho/ \zeta)^{2}$, smaller
viscosities imply longer timescales, as expected.

\subsection{Stability}\label{stability}

\begin{figure}
\plotone{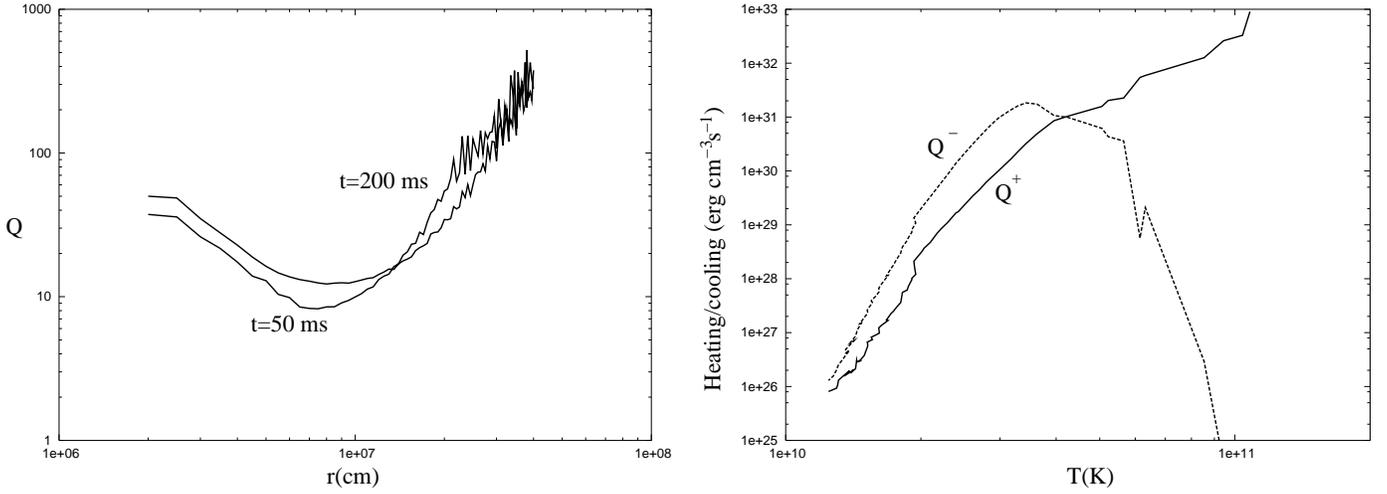}
\caption{(a) Toomre parameter $Q_{\rm T}(r)=\omega_{r} c_{s}/\pi G \Sigma$ for run
a2M at $t=50$~ms and $t=200$~ms. (b) Heating and cooling rates,
$Q^{+}$ and $Q^{-}$ as a function of the central (equatorial) disk
temperature, $T_{c}$, also for run a2M at $t=50$~ms. See text for
discussion.}
\label{Qthermalstability}
\end{figure}

Aside from convection, several general criteria for stability can be
analyzed in our case. Evidently, since we are performing dynamical
calculations, any instability that arises will quickly lead to a
change in structure. It is nevertheless instructive to consider the
corresponding conditions within the disk. We first consider the Toomre
criterion, comparing gravitational and internal energies, with
\begin{equation}
Q_{\rm T}=\frac{\omega_{r} c_{s}}{\pi G \Sigma},
\end{equation}
where $\omega_{r}$ is the local epicyclic frequency (essentially equal
in this case to the angular frequency $\Omega$), $c_{s}$ is the local
sound speed and $\Sigma$ is the surface density. We find $Q_{\rm T}>1$
in all cases throughout the calculations, and thus that the disks are
stable in this respect (Figure~\ref{Qthermalstability}a shows a
typical profile of $Q_{\rm T}[r]$).

Previous studies \citep{npk01,km02} have shown that neutrino cooled
disks are thermally unstable if radiation pressure dominates in the
flow, and stable otherwise, although the effects of neutrino opacities
were not considered. The criterion for stability can be written in
this case as
\begin{equation}
\left(\frac{d \ln Q^{+}}{d \ln T}\right)_{r} \le \left(\frac{d \ln
Q^{-}}{d \ln T}\right)_{r},
\end{equation}
where $Q^{+}$ and $Q^{-}$ are the volume heating and cooling
rates. Essentially, in order to be stable the disk must be able to get
rid of any excess internal energy generated by an increase in the
heating rate. We show in Figure~\ref{Qthermalstability}b typical
values for $Q^{+}$ and $Q^{-}$ as functions of the central
(equatorial) temperature, for run a2M at $t=50$~ms. In the optically
thin region the disks are thermally stable, because the pressure is
dominated by the contribution from free nucleons and cooling is
efficient. In the optically thick region the cooling is greatly
suppressed and the criterion would indicate that the disk becomes
thermally unstable. This is reasonable, since with optical depths
$\tau_{\nu}\approx 10-100$, essentially no direct cooling takes place,
and dissipation is not suppressed. This is why the disk is
geometrically thick in the inner regions, as already described above
(\S~\ref{structure}). The balance that gives a quasi--steady state
structure is achieved mainly through the diffusion of neutrinos, since
the diffusion timescale is shorter than the accretion timescale.

\subsection{Disk evolution}\label{evolution}

\begin{figure}
\plotone{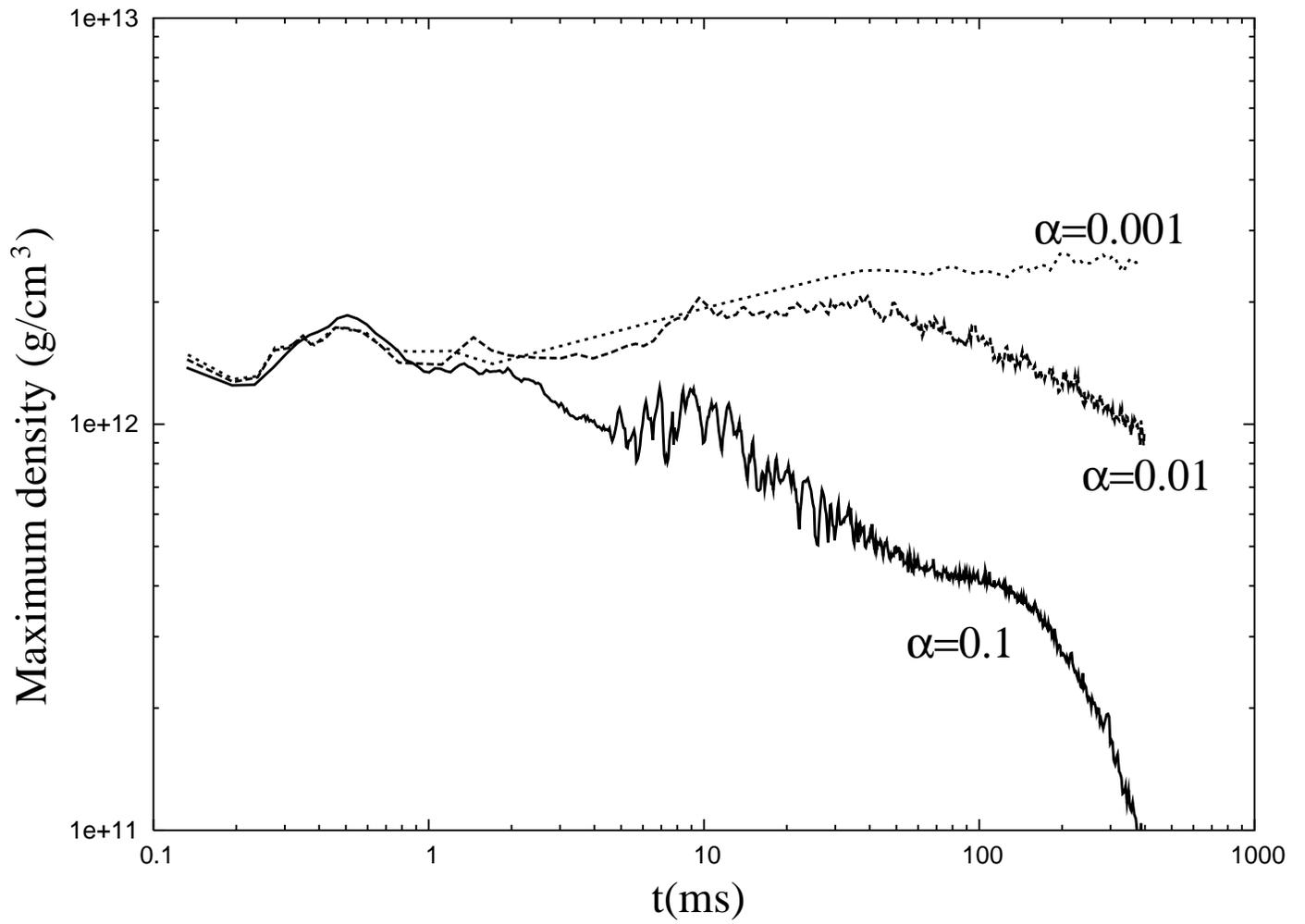}
\caption{Maximum density, $\rho_{max}$ in the accretion disk for runs
a1M, a2M and a3M.}
\label{rhomax}
\end{figure}

\begin{figure}
\plotone{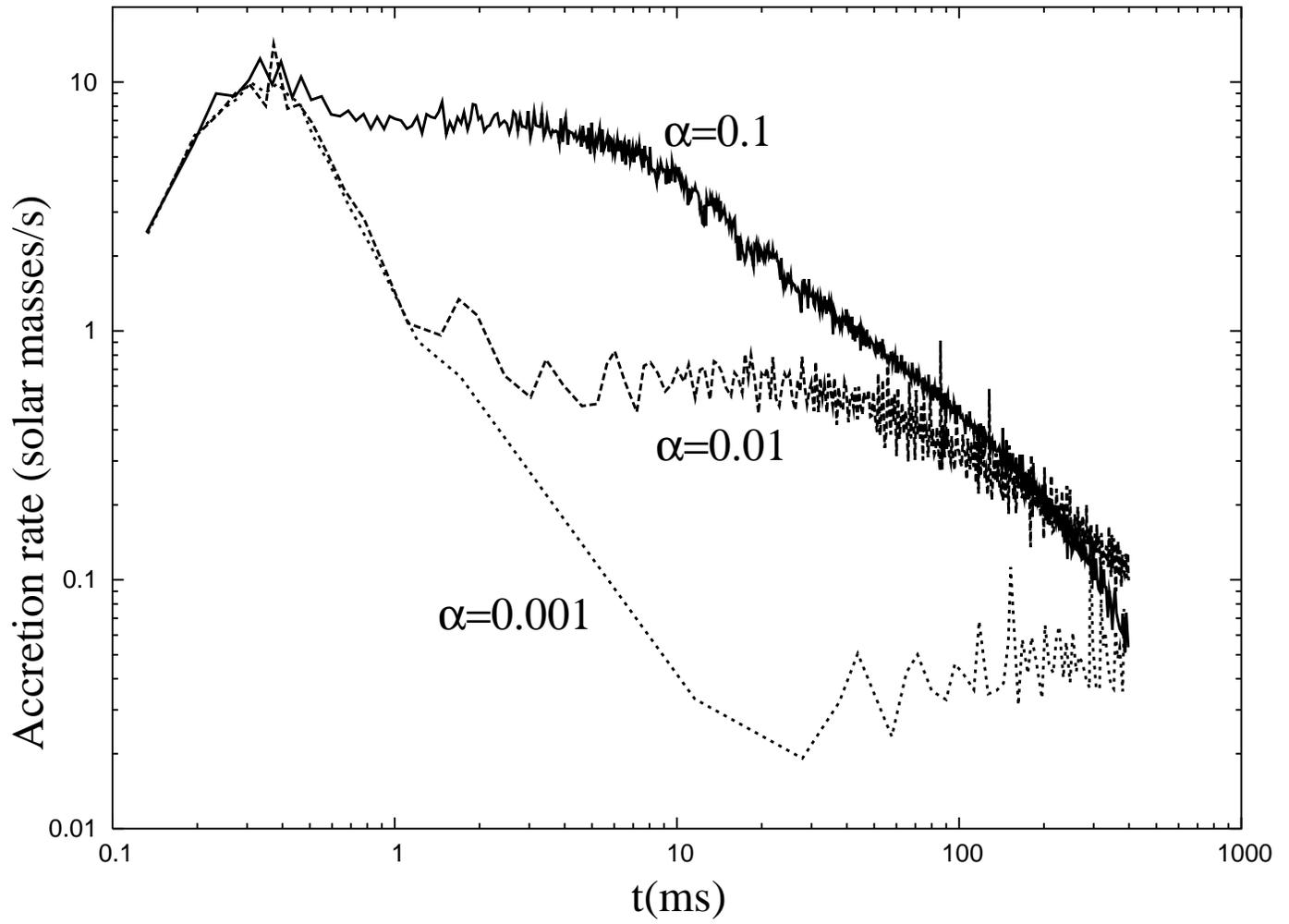}
\caption{Net accretion rate (in solar masses per second) onto the
black hole as a function of time for runs a1M, a2M, and a3M.}
\label{mdot}
\end{figure}

\begin{figure}
\plotone{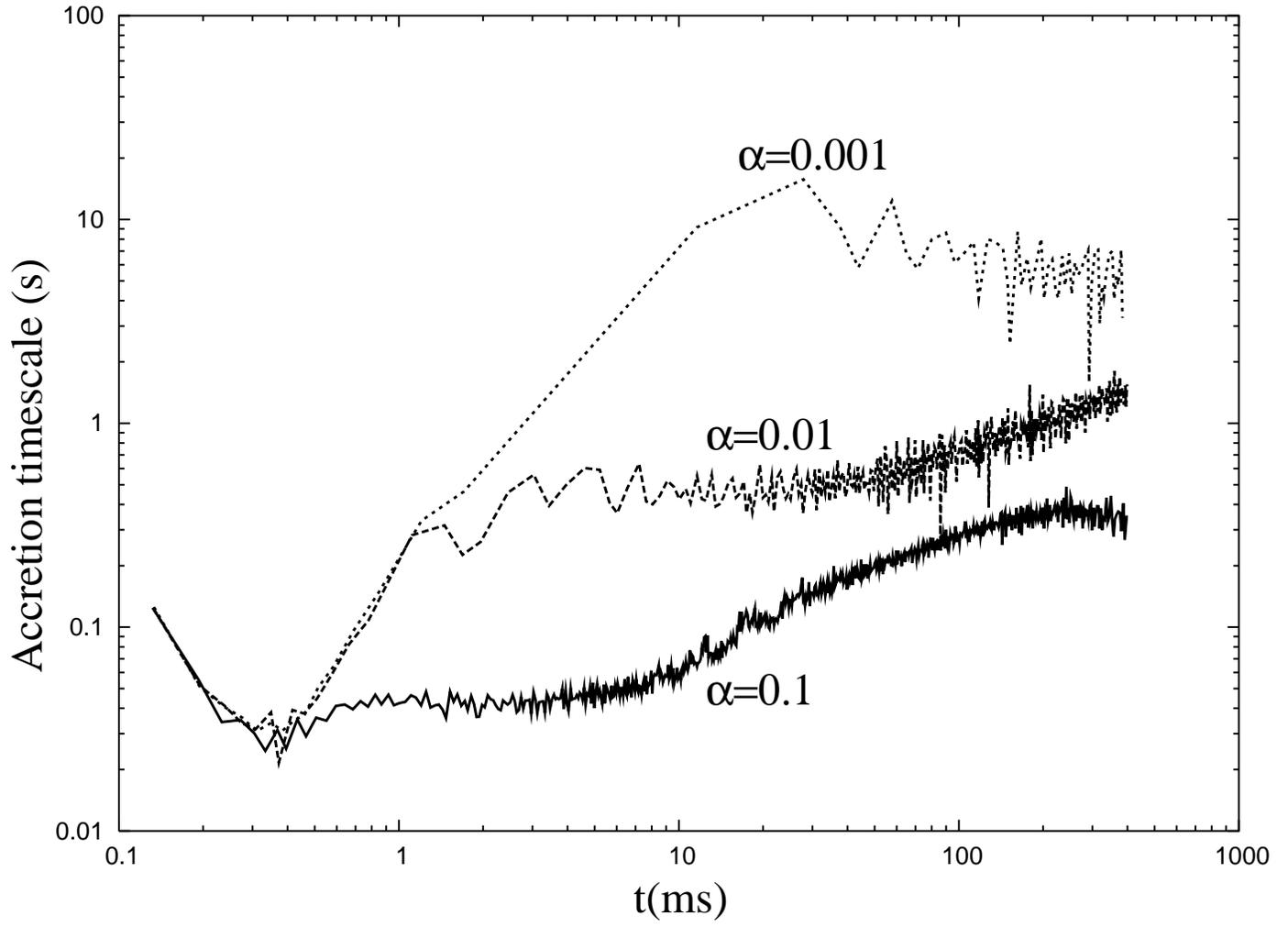}
\caption{Accretion timescale, $M_{d}/\dot{M}_{\rm BH}$ in seconds and
as a function of time for runs a1M, a2M and a3M.}
\label{tacc}
\end{figure}

\begin{figure}
\plotone{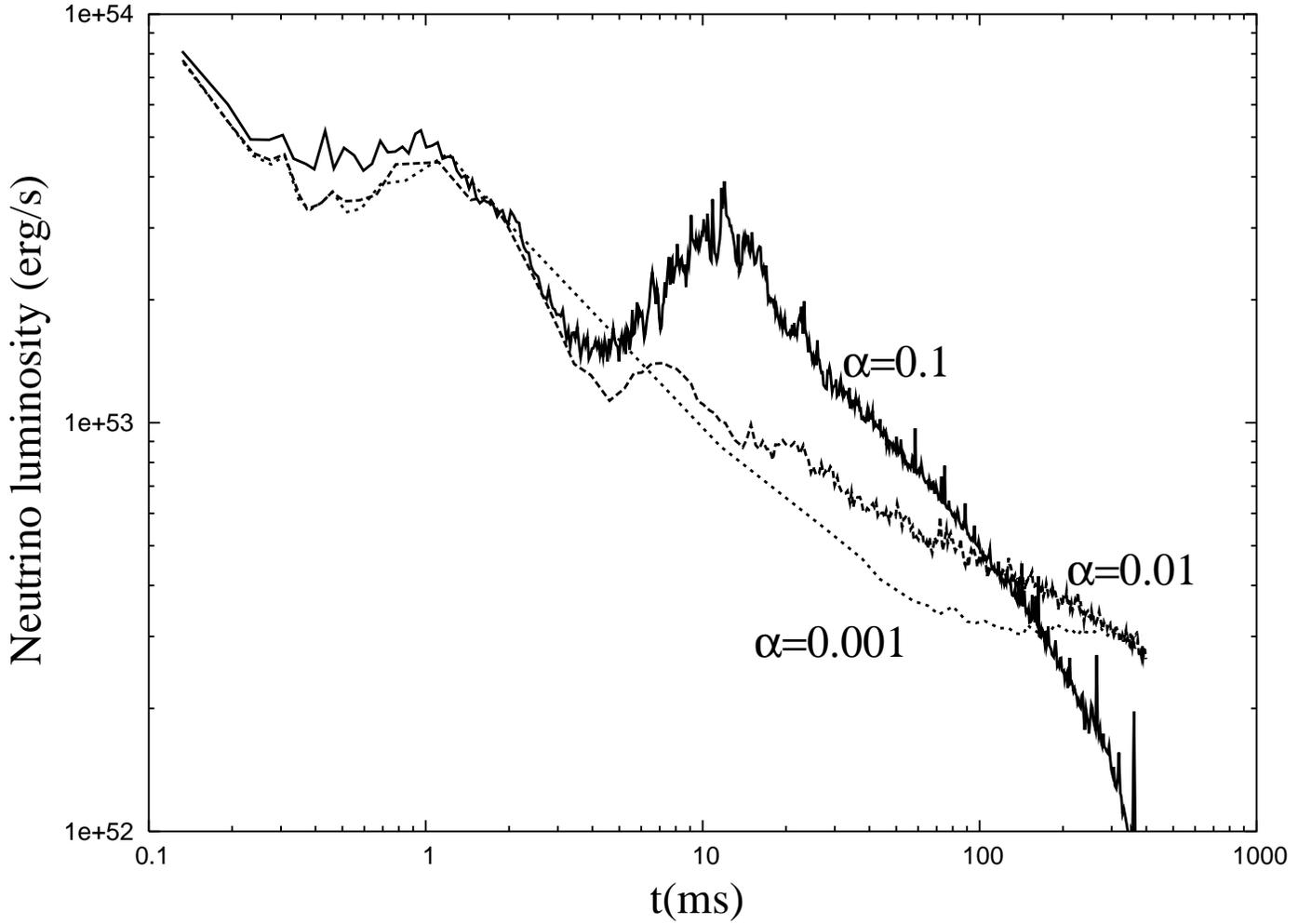}
\caption{Neutrino luminosity $L_{\nu}$, computed from
equation~(\ref{eq:nulum}), for runs a1M, a2M and a3M. Note the fairly
flat curves and rise, for run a1M at early times, when the disk is
optically thick, and the transition to the optically thin regime at a
later time. The curves for runs a2M and a3M exhibit a break
approximately on the cooling timescale, $t_{\rm cool}$. }
\label{lum}
\end{figure}

The evolution of the disk on long timescales is determined by the
balance between two competing effects: on one hand, viscosity
transports angular momentum outwards, matter accretes and the disk
drains into the black hole on an accretion timescale $t_{\rm acc}$. The
trend in this respect is towards lower densities and temperatures. On
the other hand, cooling reduces pressure support and leads to vertical
compression, increasing the density. The internal energy of the fluid
is released on a cooling timescale, $t_{\rm cool}$. The presence of an
optically thick region in the center of the disk limits the luminosity
(dominated by electron and positron capture onto free nucleons) to
$\approx {\rm few} \times 10^{53}$erg~s$^{-1}$, and the initial
internal energy is $E_{\rm int}\approx 10^{52}$~erg, so $t_{\rm cool} \simeq
0.1 $~s. If $t_{\rm acc} > t_{\rm cool}$, the disk will cool before its mass
or internal energy reservoir is significantly affected by accretion
onto the black hole. The maximum density (shown in Figure~\ref{rhomax}
for runs a1M, a2M and a3M) actually increases slightly due to vertical
compression, and subsequently drops once mass loss through accretion
dominates (the $\simeq 10$~ms delay at the start, during which it is
approximately constant, is simply the sound crossing time across the
optically thick region of the disk). The accretion rate onto the black
hole, $\dot{M}_{\rm BH}$, the accretion timescale
$t_{\rm acc}=\dot{M}_{\rm BH}/M_{\rm disk}$ and the total neutrino luminosity
$L_{\nu}$ are shown in Figures~\ref{mdot}, \ref{tacc} and
\ref{lum}. They all show the same qualitative behavior, remaining
fairly constant (or changing slowly) for an accretion timescale, and
abruptly switching thereafter. The accretion timescales are
approximately 0.5~s and 5~s for $\alpha=0.01, 0.001$ respectively.

For high viscosity, $\alpha=0.1$, the transport of angular momentum is
vigorous, and the black hole quickly accretes a substantial amount of
mass ( $0.16~M_{\sun}$ within the first 100~ms). The accretion
timescale is $t_{\rm acc} \simeq 50$~ms. The circulation pattern consists
of large--scale eddies, with $H \simeq r$. In fact there is
essentially one large eddy on each side of the equatorial plane, with
mass inflow along the surface of the disk, and an equatorial
outflow. Part of the outflowing gas moves away from the equator and
reverses direction close to the surface of the disk, contributing to
the inflow.  For an intermediate viscosity, $\alpha=0.01$, the
intensity of the circulations is smaller, but also, the eddies are
smaller, with several of them clearly occurring in the disk at
once. The transport of angular momentum being less vigorous, the
accretion rate is substantially smaller than in the previous
case. Finally, for a yet lower viscosity, $\alpha=0.001$, the trend
continues, and the eddies become smaller still. In these last two
cases, angular momentum transport is so low that very high densities
are maintained in the central regions of the disk for a large number
of dynamical times, contrary to what is seen in the high viscosity
case. A simple way to quantify how much of the mass flow is actually
making it to the central black hole is to measure the fraction of mass
at any given radius that is moving inwards,
$|\dot{M}_{\rm in}|/(|\dot{M}_{\rm in}|+|\dot{M}_{\rm out}|)$, where we have
divided the mass flow rate into two height--integrated parts, one with
$v_{r}<0$ and another with $v_{r}>0$. For example, for run a2M at
$t=100$~ms and $r=r_{*}$ it is approximately 1/3. The ratio tends to
unity only in the innermost regions of the disk, and shows the true
black hole mass accretion rate, plotted in Figure~\ref{mdot} for the
same cases.

We now turn our attention to the neutrinos, which are the main source
of cooling. The results in this case are markedly different from what
we initially found \citep{lrr02}, simply because of the new equation
of state, the more realistic cooling rates and the approximate
computation of opacities. They are in general agreement with the
preliminary results we presented before \citep{lrrp04}, which used a
less detailed equation of state than the one shown here.

The optically thick region is present in every disk at the start of
the calculation. As already mentioned, this has two important effects:
enhancing the pressure and suppressing the neutrino emission. This
reflects upon the total luminosity, since the suppression occurs
precisely in the hottest regions, where most of the energy would
otherwise be released.  The most important qualitative difference
between the runs presented here is that for a high viscosity
($\alpha=0.1$, runs a1M and a1m), the disk is drained of mass so fast
that it has no chance to cool ($t_{\rm acc}<t_{\rm cool}$) and release
most of its internal energy. It is in fact advected into the black
hole. Moreover, the optically thick region disappears entirely from
the disk (see Figure~\ref{compa0.01a00.1t100t200}) by $t=40$~ms. The
emission is then no longer suppressed (see the contours of $\dot{q}$
in the same figure), and the disk radiates at the maximum possible
cooling rate. The thermal energy content of the disk is so large, that
as it thins, the luminosity actually increases briefly around
$t=10$~ms before dropping again. The drop at late times follows an
approximate power law, with $L_{\nu} \propto t^{-1}$. In this interval
the energy release comes from a combination of residual internal
energy and viscous dissipation within the disk. For lower values of
the viscosity, $\alpha \leq 0.01$, the central regions of the disk
remain optically thick throughout the calculations. Note that reducing
the disk mass by a factor of five (as was done for runs a1m, a2m, a3m)
does not affect these overall conclusions. The fluid is simply
compressed into a smaller volume, and thus the densities and
temperatures that are reached are similar than for the high mass
runs. For run a3M the neutrino luminosity is practically constant at
$3 \times 10^{52}$~erg~s$^{-1}$ for $t\ge
100$~ms. Table~\ref{table:evol} summarizes the typical disk mass,
energy density, accretion rate, luminosity, and duration and
energetics of neutrino emission for all runs.

\section{Summary, conclusions and astrophysical implications}\label{ccl}

\subsection{Summary}
We have performed two-dimensional hydrodynamical simulations of
accretion disks in the regime of hypercritical accretion, where
neutrino emission is the main cooling agent. The disks are assumed to
be present around a stellar--mass black hole, and are evolved for a
few hundred dynamical timescales. We have paid particular attention to
the relevant microphysical processes under the conditions at hand, and
used a detailed equation of state which includes an ideal gas of
$\alpha$ particles and free baryons in nuclear statistical equilibrium
and a relativistic Fermi gas of arbitrary degeneracy. The composition
of the fluid is determined by weak interactions. The density and
temperature are such that the inner regions of the disk become opaque
to neutrinos, and this is taken into account in a simple
approximation.

\subsection{Conclusions}

Our main conclusions can be summarized as follows:
\begin{itemize}
\item Once the fluid becomes photodisintegrated into free nucleons,
neutronization becomes important and lowers the electron fraction
substantially below 1/2, with the electron fraction $Y_{e}$ reaching
$\approx 0.05$ at its minimum.  This value, however, does not occur in
the immediate vicinity of the black hole, but rather at the transition
radius where the fluid becomes optically thick to its own neutrino
emission, $r_{*}\approx 10^{7}$~cm. At smaller radii, the electron
fraction rises again, reaching $\approx 0.1$ close to the horizon.
\item Neutrino trapping produces a change in composition and an
inversion in the electron fraction. The associated negative gradient
in $Y_{e}(r)$ induces convective motions in the optically thick region
of the disk. This is analogous to what presumably occurs following
core collapse, in a proto--neutron star and its surrounding
envelope. Due to the radial flows induced by viscosity, convection is
unable to suppress this composition gradient. It would appear, however,
that the entropy per baryon in the optically thick region is very
close to being constant, with $s/k\approx 6$.
\item The spatial structure of the disk is characterized by the 
transition radius $r_{*}$ where the material becomes optically thick.
For $r>r_{*}$ the disk cools efficiently, whereas for $r<r_{*}$ 
the emission is suppressed and the fluid is unable to cool 
directly (although it does so on a neutrino diffusion timescale). 
This leads to larger pressures and a more moderate rise in density. 
\item The temporal evolution of the disk is determined by the balance
between accretion and neutrino emission. For low viscosities ($\alpha
\le 0.01$), the disk is able to cool in a quasi steady state and
radiate its internal energy reservoir.  This lasts for approximately
0.1-0.4~s, with $L_{\nu}\approx 10^{53}$~erg~s$^{-1}$. Thereafter the
typical luminosity and density quickly decay. For large viscosities
($\alpha \simeq 0.1$) the disk is drained of mass on an accretion
timescale which is shorter than the cooling timescale, and the
internal energy of the disk is essentially advected into the black
hole. An interesting result in this case is that as the disk becomes
transparent before being engulfed by the hole, it undergoes a
re-brightening, as some of the stored energy escapes.
\item The total energy output in neutrinos is $E_{\nu} \approx
10^{52}$~erg, over a timescale of $\approx 200$~ms. The typical
accretion rates are $\approx 0.1 M_{\sun}$~s$^{-1}$, and neutrino
energies are $\approx 8$~MeV at $r_{*}$. Energy densities in the inner
regions of the disk are $\approx 10^{31}$erg~cm$^{-3}$.
\end{itemize}

\subsection{Discussion}
There are two main ingredients in the results presented here that
contrast with those available previously in the literature concerning
the steady state structure of neutrino cooled accretion disks. The
first is the ability to dynamically model the evolution of the system
for hundreds of dynamical timescales, taking the previous history of
the fluid into consideration through the choice of initial
conditions. This allows us to consider the energetics on more relevant
timescales (cooling, viscous) than the dynamical one accessible in
three dimensional studies. The second is the realization that the
structure of the disks is affected qualitatively by the presence of an
optically thick region at high densities. In this sense the situation
is similar to that encountered following massive core collapse, where
convection occurs \citep[recent work assessing the relative importance
of the MRI, neutrino and convection driven viscosity in rotating
collapsing cores has been reported by][we comment further on this
issue in \S\ref{magnetic}]{awml03,tqb05}.

Clearly there is room for improvement in the results presented
here. To begin with, our expressions for the effect of a finite
optical depth on cooling and pressure are too simple, and attempt only
to capture the essential physical behavior of the system. There is an
important region of the disk where the optical depth is not large
enough to consider the diffusion approximation, and where more
detailed transport effects ought to be considered. We have not
separated the neutrino variables into three species, which would be
more rigorous \citep[e.g.,][have noted that in such flows, the
electron neutrinos $\nu_{e}$ might be preferentially absorbed with
respect to electron anti--neutrinos $\overline{\nu}_{e}$, thus
affecting the neutrino annihilation luminosity above the surface of
the disk in an important way]{yua04}. We have only considered the
effects of coherent scattering off nucleons and $\alpha$ particles for
opacity purposes. This specifically ignores absorptive scattering,
which would: (i) lead to a modification of the neutrino emergent
spectrum; (ii) produce heating of the fluid. The latter may be
particularly important concerning the driving of powerful winds off
the surface of the disk, and needs to be addressed more carefully in
the context of GRBs. As in previous work, we have chosen to maintain
the Newtonian expression for the potential well of the central black
hole for ease of comparison to previous results (this will be
addressed in future work).

\subsubsection{Gamma-Ray Bursts}
\label{GRBs}
The sudden release of gravitational binding energy of a neutron star
is easily sufficient to power a GRB.  The minimum energy requirement
is $10^{51} (\Delta \Omega/4\pi)$ erg if the burst is beamed into a
solid angle $\Delta \Omega$. As discussed in \S\ref{intro}, a familiar
possibility, especially for bursts belonging to the short duration
category, is the merger of a neutron star - black hole, or double
neutron star binary by emission of gravity waves, which, as
illustrated here, is likely to generate a black hole surrounded by a
lower mass accretion disk. How is the available rotational and
gravitational energy converted into an outflowing relativistic plasma?
A straightforward way is that some of the energy released as thermal
neutrinos is reconverted, via collisions outside the disk, into
electron-positron pairs or photons. The neutrino luminosity emitted
when disk material accretes via viscous (or magnetic) torques on a
timescale $\Delta t \sim 1$ s is roughly
\begin{equation}
L_{\nu}\sim 2 \times 10^{52} \left({M_{\rm disk} \over 0.1
M_\sun}\right)\left({\Delta t \over 1 {\rm s}}\right)^{-1}
\mbox{erg~s$^{-1}$}
\end{equation}
for a canonical radiation efficiency of 0.1. During this time, the
rate of mass supply to the central black hole is of course much
greater than the Eddington rate. Although the gas photon opacities are
large, the disk becomes sufficiently dense and hot to cool via
neutrino emission. There is in principle no difficulty in dissipating
the disk internal energy, but the problem is in allowing these
neutrinos to escape from the inflowing gas. At sufficiently low
accretion rates, $\alpha \lesssim 0.01$, we find that the energy
released by viscous dissipation is almost completely radiated away on
a timescale given by $t_{\rm cool} \approx E_{\rm int}/L_\nu \sim
0.1$~s. In contrast, for a higher mass supply, $\alpha \gtrsim 0.1$,
energy advection remains important until the entire disk becomes
optically thin. The restriction on the cooling rate imposed by high
optical depths is key because it allows the energy loss to be spread
over an extended period of time during which the neutrino luminosity
stays roughly constant. This gives a characteristic timescale for
energy extraction and may be essential for determining the duration of
neutrino-driven short GRBs \citep{lrrp04}.

Neutrinos could give rise to a relativistic pair-dominated wind if
they converted into pairs in a region of low baryon density
(e.g. along the rotation axis, away from the equatorial plane of the
disk). The $\nu\bar\nu \to {\rm e}^+ {\rm e}^-$ process can tap the
thermal energy of the torus produced by viscous dissipation. For this
mechanism to be efficient, the neutrinos must escape before being
advected into the hole; on the other hand, the efficiency of
conversion into pairs (which scales with the square of the neutrino
density) is low if the neutrino production is too gradual.  Typical
estimates suggest a lower bound\footnote{This estimate, however,
assumes that the entire surface area emits close to a single
temperature black--body. It should be noted that if the dissipation
takes place in a corona--like environment, the efficiency may be
significantly larger \citep{rrs05}.} of $L_{\nu\bar\nu} \sim 10^{-3}
L_{\nu}$ \cite[e.g][]{rrr03}. If the pair-dominated plasma were
collimated into a solid angle $\Delta \Omega$ then of course the
apparent isotropized energy would be larger by a factor $(4\pi/\Delta
\Omega)$, but unless $\Delta \Omega$ is $\leq 10^{-2}$ this may fail
to satisfy the apparent isotropized energy of $10^{52}$ ergs implied
by a redshift $z=1$ for short GRBs.

One attractive mechanism for extracting energy that could circumvent
the above efficiency problem is a relativistic magneto hydrodynamic
(MHD) wind \citep{u92,t94}. Such a wind carries both bulk kinetic
energy and ordered Poynting flux, and it possible that gamma-ray
production occurs mainly at large distances from the source
\citep{dt92,u94,t94}. A rapidly rotating neutron star (or disk)
releases energy via magnetic torques at the rate $L_{\rm mag} \sim
10^{49} B_{15}^2P^{-4}_{-3} R_6^6\;{\rm erg\;s^{-1}}$, where
$P=10^{-3}P_{-3}$ s is the spin period, and $B=10^{15}B_{15}$ G is the
strength of the poloidal field at a radius $R=10^6R_6$ cm. The last
stable orbit for a Schwarzschild hole lies at a coordinate distance
$R=6GM/c^2=9(M/M_\sun)$ km, to be compared with
$R=GM/c^2=3/2(M/M_\sun)$ km for an extremal Kerr hole. Thus the
massive neutron disk surrounding a Schwarzschild black hole of
approximately $2 M_{\sun}$ should emit a spin-down luminosity
comparable to that emitted by a millisecond neutron star. A similar
MHD outflow would result if angular momentum were extracted from a
central Kerr hole via electromagnetic torques \citep{bz77}. The field
required to produce $L_{\rm mag} \geq 10^{51}\;{\rm erg\;s^{-1}}$ is
colossal, and may be provided by a helical dynamo operating in hot,
convective nuclear matter with a millisecond period
\citep[][]{dt92}. A dipole field of the order of $10^{15}$ G appears
weak compared to the strongest field that can in principle be
generated by differential rotation ($\sim 10^{17}[P/1\;{\rm
ms}]^{-1}\;{\rm G}$), or by convection ($\sim 10^{16}\;{\rm G}$),
although how this may come about in detail is not resolved.  We
examine in more detail the possible generation of strong magnetic
fields below in \S\ref{magnetic}.

Computer simulations of compact object mergers and black hole
formation can address the fate of the bulk of the matter, but there
are some key questions that they cannot yet tackle. In particular,
high resolution of the outer layers is required because even a tiny
mass fraction of baryons loading down the outflow severely limits the
attainable Lorentz factor - for instance a Poynting flux of $10^{53}$
erg could not accelerate an outflow to $\Gamma \gtrsim 100$ if it had
to drag more than $\sim 10^{-4}$ solar masses of baryons with it. One
further effect renders the computational task of simulating jet
formation even more challenging. This stems from the likelihood that
the high neutrino fluxes ablate baryonic material from the surface of
the disk at a rate \citep{qw96}
\begin{equation} 
\dot{M}_\eta \sim 5 \times 10^{-4}\left({L_\nu \over 10^{52}\;{\rm
erg\;s^{-1}}}\right)^{5/3} M_\sun {\rm s^{-1}}.
\label{ablation}
\end{equation}
A rest mass flux $\dot{M}_\eta$ limits the bulk Lorentz factor of the
wind to 
\begin{equation} 
\Gamma_\eta={L_{\rm mag} \over \dot{M}_\eta c^2}=10 \left({L_{\rm mag}
\over 10^{52}\;{\rm erg\;s^{-1}}}\right) \left({\dot{M}_\eta \over 5
\times 10^{-4} M_\sun {\rm s^{-1}}}\right)^{-1}.
\end{equation}
If one assumes that the external poloidal field strength is limited by
the vigor of the convective motions, then the spin-down luminosity
scales with neutrino flux as $L_{\rm mag}\propto B^2\propto v_{\rm
con}^2\propto L_\nu^{2/3}$, where $v_{\rm con}$ is the convective
velocity. The ablation rate given in equation (\ref{ablation}) then
indicates that the limiting bulk Lorentz factor $\Gamma_\eta$ of the
wind decreases as $L_\nu^{-1}$. Thus the burst luminosity emitted by a
magnetized neutrino cooled disk may be self-limiting. The mass loss
would, however, be suppressed if the relativistic wind were collimated
into a jet. This suggests that centrifugally driven mass loss will be
heaviest in the outer parts of the disk, and that a detectable burst
may be emitted only within a certain solid angle centered on the
rotation axis \citep[see e.g.,][]{rrr03}.

\subsubsection{Generation of Strong Magnetic Fields}
\label{magnetic}
 
There is also the question of magnetic fields, which we have not
included, but should obviously be considered. The field in a standard
disk is probably responsible for viscous stresses and dissipation,
through the MRI. In this respect the current scenario should exhibit
these characteristics. The MRI operates on an orbital timescale, and
so the field would grow in a few tens of milliseconds.  It may also be
amplified by the convective motions described above,
\S~\ref{nuconvection}. The saturation value for the field can be
naively estimated as that at which its energy density is in
equipartition with the gas, $B^{2}/8\pi \approx \rho c_{s}^{2}$, or
when the Alfven speed, $v_{A}=B/\sqrt{4\pi \rho}$ is comparable with
the azimuthal velocity, $v_{\phi}$. This gives $B\approx
10^{16}$~G. It is not clear at all, however, that the field amplitude
will reach such high levels, because the magnetic Reynolds number is
far beyond its critical value (where diffusion balances dynamo--driven
growth), and amplification can lead to field expulsion from the
convective region, thereby destroying the dynamo
\citep{retal02}. Precious little is known about the growth of magnetic
fields at such overcritical levels \citep{rg05}, and a definitive
answer will require the self--consistent inclusion of full MHD into
the problem at hand (but with a level of resolution which may be well
above present computational capabilities). It is not clear either that
the magnetic shearing instability can generate a mean poloidal field
as strong as $B\approx 10^{16}$~G, since to first order it does not
amplify the total magnetic flux.  The non-linear evolution of the
instability depends sensitively on details of magnetic reconnection,
and it has indeed been suggested that this can smooth reversals in the
field on very small scales, pushing the dominant growing mode to much
larger scales \citep{gx94}.

It is certainly possible, as shown here, that compact binary mergers
do form a neutron disk that is hot enough to be optically thick to
neutrinos, and convective instability is a direct consequence of the
hot nuclear equation of state. A neutron disk is likely to be
convective if the accretion luminosity exceeds $10^{50}-10^{51}\;{\rm
erg\;s^{-1}}$. Note that even if the accretion luminosity is lower, a
hot, massive disk \citep[such as those forming in collapsars,][]{w93}
would undergo a brief period of convection as a result of secular
cooling (notice that convection is driven by secular neutrino cooling,
whereas the MRI is powered by a release of shear kinetic energy). If
the dense matter rotates roughly at the local Keplerian angular
velocity, $\Omega \simeq (GM_{\rm BH}/r^{3})^{1/2}$, then $L_{\rm
mag}$ is approximately independent of radius, and the required
poloidal field for a given luminosity is $B_{15} \approx L_{\rm mag,
50}^{1/2} (M_{\rm BH}/M_\sun)^{-1}$. If a period of convection is a
necessary step in the formation of a strong, large-scale poloidal
field, an acceptable model thus requires that the surrounding torus
should not completely drain into the hole on too short of a
timescale. Whether a torus of given mass survives clearly depends on
its thickness and stratification, which in turn depends on internal
viscous dissipation and neutrino cooling.

A large amount of differential rotation (as may occur in newborn
neutron stars or those in X--ray binaries, and is definitely the case
in toroidal structures supported mainly by centrifugal forces),
combined with short periods, may produce substantial magnetic field
amplification \citep{kr98,s99}. The energy transferred to the magnetic
field is released in episodic outbursts when the buoyancy force allows
the field to rise to the surface of the star or disk. The
amplification of a magnetic field to such strong values would clearly
have important consequences on the evolution and time variability of
the disk and its energy output. It would probably lead to strong
flaring and reconnection events accompanied by the release of large
amounts of energy, if the growth time for field amplification, $t_{\rm
B}$, is shorter than the accretion timescale, $t_{\rm acc}$ (otherwise
the disk would drain of matter before the field had the chance to
reach large values; in this case, the survival of a massive, rapidly
rotating neutron star as the end-point of binary NS merger might be
preferred over the prompt formation of a BH). An effective helical
dynamo of the $\alpha-\Omega$ type should be favored by a low
effective viscosity, $\alpha$, because, as stated in \S
\ref{nuconvection}, the overturn time\footnote{Recall that the
resulting convective motions in a proto-neutron star are extremely
vigorous, with an overturn time of $\sim 1$ ms.} is $t_{\rm
con}\approx 20~$ms (this applies only if the disk is not fed with
matter externally for a time longer than $t_{\rm acc}$, otherwise
convection would also be able to amplify the magnetic field).

\subsubsection{Nucleosynthesis}

Core collapse SNe and compact object mergers are natural astrophysical
sites for the production of heavy elements
\citep{ls74,mb97,fetal99,retal99,lee01}. In particular,
nucleosynthesis in neutrino--driven winds is an issue that may be
relevant for iron--group elements, as well as for heavier nuclei
through the r--process \citep{wh92}. Initial investigations into this
matter \citep{qw96} determined that the entropy in the outflow arising
from a newborn neutron star was probably too low to give rise to the
r--process efficiently. More recently, this problem has been addressed
again in the specific case of collapsar or post--merger accretion
disks, based on the results of analytical calculations of neutrino
cooled disks in one dimension \citep{pth04}. If the wind consists of a
uniform outflow driven from the surface of the disk, the entropy is
too low, and essentially only iron--group elements are synthesized, in
agreement with the earlier results. However, their results also
indicate that in a bubble-type outflow (against a background steady
wind), where episodic expulsion of material from the inner regions
takes place, material may be ejected from the disk and preserve its
low electron fraction, thus allowing the r-process to occur. The
convective motions reported here, occurring in the optically thick
portion of the disk, would represent one way such cells could be
transported to the disk surface if they can move fast enough to
preserve the neutron excess. Since the existence of a convection
region is dependent upon the densities reached in the inner disk (so
that it becomes opaque), the synthesized nuclei (iron group
vs. heavier, r--process elements) could be a reflection of its
absence/presence.

\acknowledgments We have benefited from many useful discussions and
correspondence with U. Geppert, N. Itoh, K. Kohri, P. Kumar,
A. MacFadyen, P. M\'{e}sz\'{a}ros, M. Prakash, M. Rees, T. Thompson,
S. Woosley and A. Socrates.  Financial support for this work was
provided in part by CONACyT 36632E (WHL, DP) and by NASA through a
Chandra Postdoctoral Fellowship award PF3-40028 (ERR). Part of this
work was done during visits to the Institute for Advanced Study (WHL)
and Instituto de Astronom\'{\i}a, UNAM (ERR), whose hospitality is
gratefully acknowledged. We thank the anonymous referee for helpful
comments and suggestions on the initial manuscript.

\clearpage

\clearpage

\clearpage

\clearpage

\clearpage 

\clearpage

\begin{figure}
\plotone{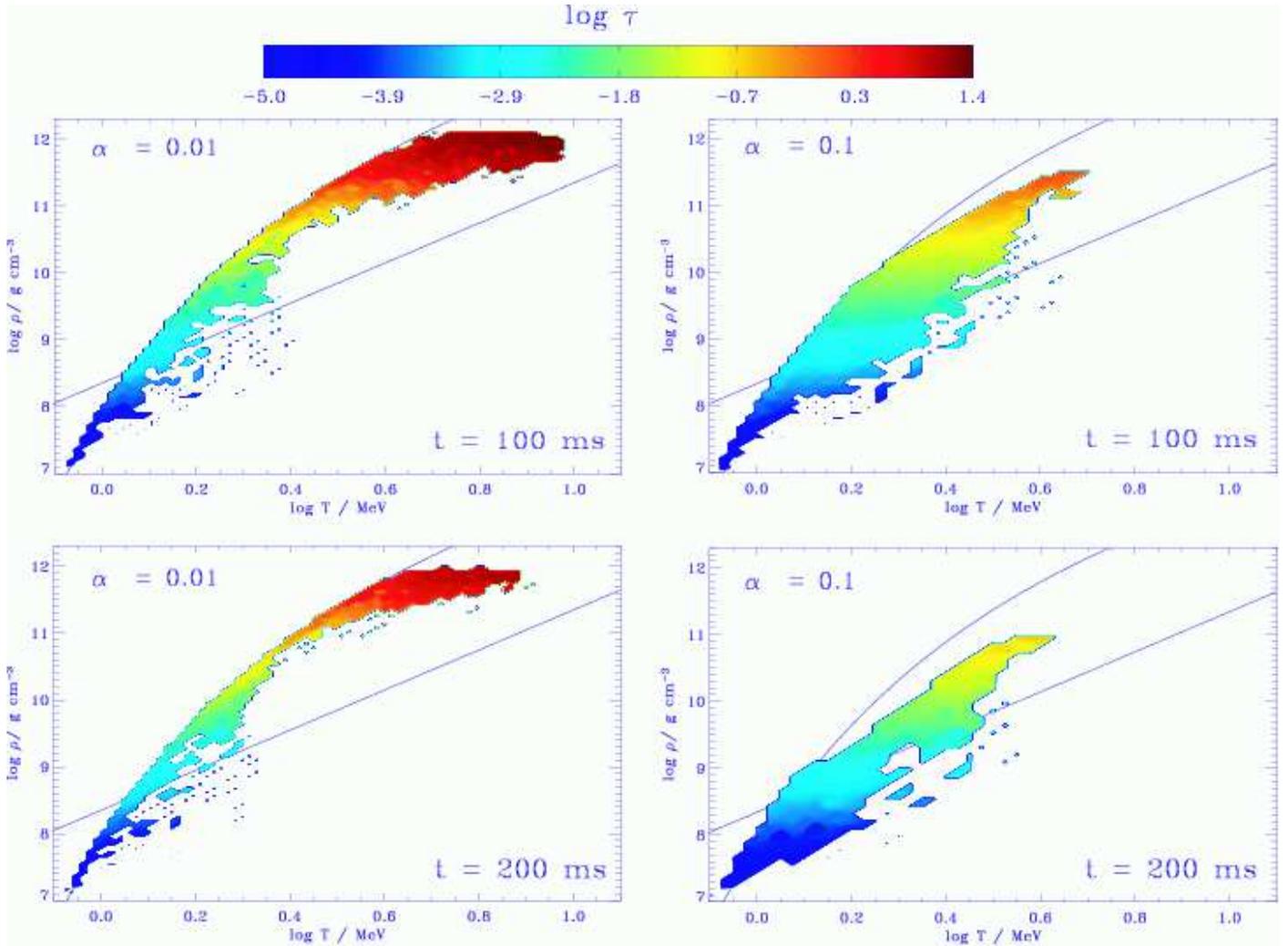}
\caption{Same as Figure~\ref{rtye}, but color coded according to
optical depth.}
\label{rttau}
\end{figure}

\clearpage

\clearpage 

\clearpage

\clearpage 

\clearpage

\clearpage

\clearpage

\begin{deluxetable}{clccc}
\tablecaption{Initial conditions for the accretion
disks.\label{table:ICs}} \tablewidth{0pt} 
\tablehead{ \colhead{Run\tablenotemark{c}} 
& \colhead{$\alpha$} & \colhead{$M_{\rm
BH}/M_{\sun}$\tablenotemark{a}} &
\colhead{$M_{disk}/M_{\sun}$\tablenotemark{a}} &
\colhead{$N$\tablenotemark{b}} }
\startdata 
aIM & 0.0 & 3.85 & 0.308 & 19,772 \\
a1M & 0.1 & 3.85 & 0.308 & 19,772 \\
a2M & 0.01 & 3.85 & 0.308 & 19,772 \\
a3M & 0.001 & 3.85 & 0.308 & 19,772 \\
a1m & 0.1 & 3.85 & 0.062 & 19,772 \\
a2m & 0.01 & 3.85 & 0.062 & 19,772 \\
a3m & 0.001 & 3.85 & 0.062 & 19,772 \\
\enddata 
\tablenotetext{a}{Values at the start of the two--dimensional
calculation of the accretion disk evolution.}
\tablenotetext{b}{Number of SPH particles at the start of the
two--dimensional calculation.}
\tablenotetext{c}{In the run label, the number refers to the value of
$\alpha$ (with ``I'' standing for inviscid, with $\alpha=0$) and the
letter to high (M) or low (m) disk mass.}
\end{deluxetable}

\begin{deluxetable}{cccccccc}
\tabletypesize{\scriptsize}
\tablecaption{Accretion disk parameters during the dynamical evolution.
\label{table:evol}}
\tablewidth{0pt}
\tablehead{
\colhead{Run} & \colhead{$\dot{M}/(M_{\sun}/s)$} &
\colhead{$\rho
c_{s}^{2}/(\mbox{erg}~\mbox{cm}^{-3})$} &
\colhead{$L_{\nu}(\mbox{erg}~\mbox{s}^{-1})$} &
\colhead{$E_{\nu}(\mbox{erg})$\tablenotemark{a}} &
\colhead{$E_{\nu}(\mbox{MeV})$\tablenotemark{d}} &
\colhead{$T_{\nu,1/2}(\mbox{ms})$\tablenotemark{b}} &
\colhead{$M_{disk}/M_{\sun}$\tablenotemark{a}} } \startdata
aIM & 0 & 6$\cdot10^{30}$ & 2$\cdot10^{52}$ & 1.5$\cdot10^{52}$
& 8 & $>$140\tablenotemark{c} & 0.308 \\
a1M & 7 & 2$\cdot10^{31}$ & 2$\cdot10^{53}$ & 2$\cdot10^{52}$
& 8 & 60 & 0.098 \\
a2M & 0.7 & 2$\cdot10^{31}$ & 6$\cdot10^{52}$ & 2$\cdot10^{52}$
& 8 & 150 & 0.2 \\
a3M & 0.05 & 1.6$\cdot10^{31}$ & 2.$\cdot10^{52}$ & 1.55$\cdot10^{52}$
& 8 & $>$140\tablenotemark{c} & 0.287 \\
a1m & 0.7 & 4$\cdot10^{30}$ & 2$\cdot10^{52}$ & 5.4$\cdot10^{51}$
& 8 & 25 & 0.022 \\
a2m & 0.05 & 3$\cdot10^{30}$ & 1$\cdot10^{52}$ & 5.5$\cdot10^{51}$ 
& 8 & 75 & 0.045 \\
a3m & 0.005 & 3$\cdot10^{30}$ & 1.5$\cdot10^{52}$ & 7$\cdot10^{51}$ 
& 8 & $>$160\tablenotemark{c} & 0.059 \\
\enddata
\tablenotetext{a}{Values are given at $t=0.2$~s.}
\tablenotetext{b}{This is the duration of the neutrino emission, 
measured as the time needed for the disk to release one half of $E_{\nu}$.}
\tablenotetext{c}{This is a lower limit, since for $\alpha=0.001$ the
disk is still in a quasi--steady state at $t=0.2$~s and has not yet
drained into the black hole.}
\tablenotetext{d}{The typical neutrino energy is given at the radius
where $\tau_{\nu}=1$.}
\end{deluxetable}

\end{document}